\documentclass[a4paper,fleqn]{cas-sc}
\usepackage[authoryear]{natbib}

\usepackage{lineno}
\usepackage{subcaption}
\usepackage{lipsum}
\usepackage[dvipsnames]{xcolor}
\usepackage{longtable}  
\usepackage{booktabs}
\usepackage{caption} 
\def\tsc#1{\csdef{#1}{\textsc{\lowercase{#1}}\xspace}}
\tsc{WGM}
\tsc{QE}
\tsc{EP}
\tsc{PMS}
\tsc{BEC}
\tsc{DE}

\begin{document}
\let\WriteBookmarks\relax
\def\floatpagepagefraction{1}
\def\textpagefraction{.001}
\shorttitle{Primitive asteroids in the Solar System}
\shortauthors{N. El-Bez-Sebastien et~al.}

\title [mode = title]{Primitive asteroids in the main belt, Cybele, and Hilda populations from Gaia DR3}                      

\author[1]{N. El-Bez-Sebastien}[type=editor,
                orcid={0009-0002-7120-4660}
]

\credit{Writing - original draft, Conceptualization, Investigation, Formal analysis, Visualization}

\affiliation[1]{organization={LIRA, Université Paris Cité, Observatoire de Paris, Université PSL, Sorbonne Université, CY Cergy Paris Université, CNRS},
                city={Meudon},
                postcode={92190}, 
                country={France},
}

\author[1,2]{S. Fornasier}
\credit{Supervision, Validation, Project administration, Conceptualization, Writing - review \& editing, Funding acquisition}
\author[1]{A. Seurat}
\credit{Visualization, Formal analysis}

\affiliation[2]{organization={Institut Universitaire de France (IUF)},
                addressline={ 1 rue Descartes}, 
                postcode={75231}, 
                city={1 PARIS CEDEX 05},
                country={France}}

\author[1,3]{A. Wargnier}
\credit{Visualization, Writing - review \& editing}

\affiliation[3]{organization={Institute of Space and Astronautical Science, Japan Aerospace Exploration Agency},
city={Sagamihara, Kanagewa},
country={Japan}}

\cortext[cor1]{Corresponding author at: LIRA, Université Paris Cité, Observatoire de Paris, Université PSL, Sorbonne Université, CY Cergy Paris Université, CNRS, Meudon, 92190, France}
\cortext[cor2]{E-mail adress: noemie.el-bez-sebastien@obspm.fr (N. El-Bez-Sebastien)}

\begin{abstract}
Primitive asteroids include C-, P-, and D-classes, known to be dark and having spectra mostly featureless. They differ in the spectral slope, which ranges from moderate values for C-types, and progressively increases in P- and D-types, the latter being the reddest. While C- and P-types are commonly observed in the asteroid main belt, D-types are commonly found further from the Sun, in the Cybele, Hilda, and Jupiter Trojans regions, and very few are reported in the main belt. 
This study aims at characterizing the abundance of primordial and red asteroids, belonging to the P-, D-, and Z-classes in the \citet{Mahlke_2022} taxonomy, in the 2-5.2 AU region using the third data release by the Gaia mission (DR3) spectral catalog, which includes more than 60000 spectrophotometric data of asteroids. We have applied the following methodology to identify primordial asteroids in the catalog: 1{)} selection of objects with signal to noise ratio (SNR) greater than 20; 2{)} albedo value less than 12\%; 3{)} chi-squared fit to automatically identify potential D-, Z-, and P-types using Bus-DeMeo and Mahlke taxonomy; 4{)} visual inspection of every spectrum to confirm the taxonomic classification.  

Referring to Mahlke taxonomy, we have found 318 new D-types across the main belt, as well as 124 Z-types, which is a considerable increase from previous studies \citep{Mahlke_2022,DeMeoDtype,oriel_2024_red,Gartrelle2021b}, and is in agreement with theoretical estimations \citep{DeMeoDtype,volkrou2016}. We look for correlations among physical and orbital parameters, and we computed the spectral slope in the visible range (0.55 -- 0.81 $\mu$m). We also have identified 265 P-types in the main belt. For the Cybele and Hilda asteroids, we characterize the taxonomic class of all the bodies with SNR higher than 20 in the Gaia catalog, for a total sample of 193 and 180 asteroids, respectively.      

For D- and Z-types in the main belt, we have found a correlation between size and semimajor axis, meaning that the D-types closer to the Sun are smaller and also brighter, which could uncover their implantation processes, such as collisions or Yarkovsky effect. No families dominated by D- or Z-types have been identified so far. These red asteroids were likely formed further from the Sun and implanted during the giant planets' migrations. 
 
For both the Cybele and Hilda, we found a bimodal spectral slope distribution as already reported in the literature. These groups are dominated by P- and D-types. Furthermore, for both populations, despite not being the most abundant group, P-types show a larger range of diameters and include the biggest bodies, hinting at a more robust material. 

We have identified 42 D- and 18 Z-asteroids having orbits with high inclination and eccentricity, and Tisserand parameter lower than 3, which could have cometary origins, in the main belt, Cybele, and mostly in the Hilda. We also compared the mean D- and Z-types spectra with those of the two Martian moons, Phobos and Deimos.  Deimos is spectrally closer to Z-types, while Phobos red unit to D-types, confirming the possibility that Mars' satellites might be captured asteroids.
\end{abstract}


\maketitle

\section{Introduction}

In Tholen and Mahlke's taxonomy, primitive asteroids include C-, P-, and D-types, which are known to be featureless or to have faint bands in the visible-near-infrared (VNIR) region. D-type asteroids are characterized by their spectral redness and low albedo, as well as being featureless in the visible spectrum \citep{1984PhDTholen, demeo_tax, Mahlke_2022}. Recently, \citet{Mahlke_2022} separated the D-type class into two distinct types: D-types and Z-types. Z-types are characterized by a significantly redder spectral slope. Both D- and Z-types are thought to be rich in organics, as confirmed by JWST observations showing that redder D-types in the Jupiter Trojan swarms exhibit organic absorption bands \citep{jwst_lucy}. D-types are rarely observed in the main belt and are typically found in the outer solar system, especially beyond 3.5 AU. They are the dominant class of Jupiter Trojans \citep{gradie1982,Fornasier2004,Fornasier2007,Fornasier2025,carvano2010, Demeo2013}. 

Thanks to the SMASS I survey, the first six D-type objects were discovered in the main belt \citep{Xu95}. Later, \citet{Bus99} detected eight of them using SMASSII data. The Sloan Digital Sky Survey (SDSS) recovered additional D-types in the main belt, including those in the inner belt, which extends from 2.0 to 2.5 AU.  However, the vast majority were found beyond 3.5 AU \citep{carvano2010, Demeo2013}.

The latter authors classified a total of 546 D-types across the Solar System, though only a few were found within the 2.0–2.5 AU range. The authors estimated that D-types comprise 1\% of the mass in the inner main belt, 2\% in the middle, and 8\% in the outer belt. Subsequent spectral observations by \citet{DeMeoDtype} confirmed only 20\% of these D-types and found that those in the inner main belt are, on average, brighter than those farther from the Sun and have small diameters (D $<$ 30 km). The authors estimated that there should be around 100 small D-types in the inner main belt.
Using the MOVIS catalog containing near-infrared colors, \citet{popescu2018a} found five potential D-types in the inner main belt (IMB). \citet{Gartrelle2021b} conducted a study on 86 D-types at different heliocentric distances, including 25 new observations and 15 objects not previously observed, including Jupiter Trojans. They found that the spectral slope increases in the visible near-infrared as the objects get closer to the Sun. 
Following the SDSS detection of D-types in the main belt, \citet{oriel_2024_red}  observed 32 of them and confirmed 50\% as D-types. The observations revealed two groups: some exhibited a steep spectral slope in the longer near-infrared wavelengths, while others showed flattening, suggesting diverse origins or compositions. \citet{Mahlke_2022} reported 82 D-types and 23 Z-types, of which only one D-type and six Z-types were in the inner main belt, four D-types and six Z-types were in the middle belt, and five D-types and three Z-types were in the outer belt. \\

Red featureless asteroids likely originated from the outer Solar System, potentially from the trans-Neptunian object (TNO) population \citep{levison2009}. This theory is supported by the discovery of extremely red asteroids in the main belt, such as 203 Pompeja and 269 Justitia, which have spectra resembling those of TNOs and are believed to have formed in the outer Solar System \citep{ast269}. \citet{levison2009} showed that the giant planet's migration produced the implantation of trans-Neptunian objects (TNOs) in the outer main belt, but failed to implant TNOs in the inner belt. \citet{DeMeoDtype} suggest that implanted TNOs crossed the 3:1 resonance via Yarkovsky force or from collisional processes. 

Additional reservoirs of primordial asteroids  are the Cybele and the Hilda populations, located in between the main belt and the Jupiter Trojans \citep{dahlgren1997, gilhutton2008, DePra2018}. The Cybele group is in the 2:1 mean motion resonance with Jupiter, between 3.27 and 3.7 AU, while Hilda asteroids are located around 4.2 AU, at the 3:2 mean motion resonance with Jupiter. Hilda asteroids are thought to share a common origin with Trojans \citep{marsset2014}. In contrast, Cybele exhibits greater taxonomic diversity, including P/D- and X-types, with a smaller proportion of D-types compared to Hilda \citep{lagerkvist2005, gilhutton2010}.

In this paper, we conduct a statistical analysis of main belt asteroids classified as D-, Z-, and P-types using the Gaia Data Release 3 (DR3) spectral catalog, following the \citet{Mahlke_2022}  classification scheme. We also analyze the Cybele and Hilda populations, all taxonomies considered.

\section{Methods}

\subsection{Gaia DR3 data}

The Gaia third data release (DR3) spectral catalog includes spectra of 60,518 asteroids \citep{galluccio:hal-04225764}. The data were acquired with the Blue and Red Photometers (BP and RP, respectively), which cover the 0.33--0.68 $\mu$m and the 0.64--1.05 $\mu$m ranges, respectively. Each spectrum is an average of the different reflectivities measured between 2014 and 2017. The spectral data in the catalog are averaged into 16 spectrophotometric points across bins of 0.044 $\mu$m. They are centered at fixed wavelengths (0.374, 0.418, 0.462, 0.506, 0.550, 0.594, 0.638, 0.682, 0.726, 0.770, 0.814, 0.858, 0.902, 0.946, 0.990, and 1.034 $\mu$m). The relative reflectance is normalized at 0.55 $\mu$m. The analysis of Gaia DR3 asteroid data is exactly the same as that reported by \citet{Fornasier2025}, who studied the Jupiter Trojans present in the Gaia spectral catalog. 

We have used the Bus-DeMeo (BDM) taxonomy \citep{demeo_tax} and the Mahlke taxonomy \citep{Mahlke_2022} to classify the asteroid spectra. The main difference between the two is that Mahlke's taxonomy uses albedo to define asteroid classification, while Bus-DeMeo's does not. Moreover, it introduces a new type of interest to us: the Z-type. Z-types are similar to D-types in that they are featureless and dark; however, they differ in that they are significantly redder. Moreover, at longer wavelengths they tend toward a more convex shape. Nonetheless, this behavior cannot be appreciated in the visible, hence we distinguish between D- and Z-types solely on the basis of the visible spectral slope. The D- and Z-class have a spectral overlapping region in the visible \citep{Mahlke_2022}. Hence a small amount of red D-types might have been classified Z-, and conversely less red Z-types might have been classified D-. Near-infrared observations would be necessary to definitely confirm a D- or Z- classification.

\begin{figure}

    \includegraphics[width=\hsize]{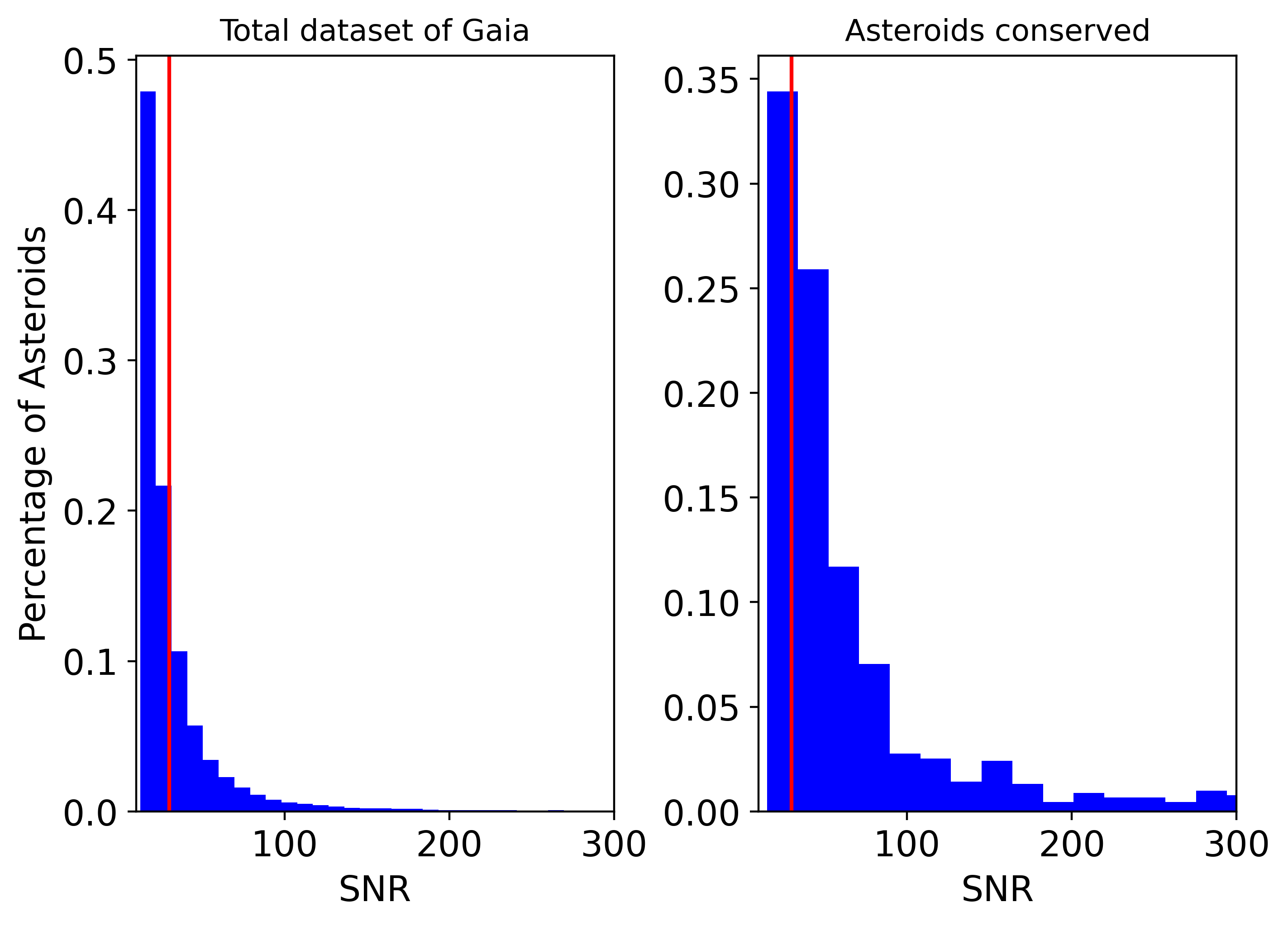}
    \caption{Distribution of the SNR for the entirety of Gaia's catalog (left panel) and for the retrieved dataset (right panel) composed of D-, Z-, and P-types. The red lines represent the SNR threshold at 20.}
    \label{fig:SNR}
\end{figure}

In the Gaia dataset, we look for primordial asteroids located in the 2--4.4 AU range (Trojans were already investigated by \citet{Fornasier2025}, we will comment on their results later).
We performed a data pre-selection based on the quality of the spectra. To do so, we computed the signal to noise ratio as: $$S/N =\frac{1}{N}\sum_{n=1}^N{R(\lambda_n)/R_{error}(\lambda_n})$$ where $R(\lambda)$ is the normalized reflectance for a given wavelength $\lambda$, N the number of points in the spectrum and $R_{error}$ the associated error reported for each Gaia reflectance. We only kept spectra that had a SNR greater than 20, which represents 59\% of the catalog, as it is considered a trustworthy threshold \citep{galinier:tel-04814227}. We have, however, been careful with objects with SNR between 20 and 30 (26.78\% of the catalog), checking individually the quality of the spectra. The SNR distribution of Gaia, and the distribution of our dataset, is shown in Fig. \ref{fig:SNR}.\\
The Gaia spectrophotometry of a given asteroid is the mean of several observations acquired at different epochs and phase angle conditions. The mean phase angle of MBA in the Gaia DR3 catalog is of about 20°, ranging from 5 up to 30° \citep{galluccio:hal-04225764}. \citet{wargnier_2025_spec} estimated the limit of phase reddening on Gaia data to be 0.01 \%/100 nm/deg. For comparison, \citet{perna2018} found a spectral phase reddening coefficient of -0.024 $\pm$ 0.027 \%/100 nm/deg for Near-Earth D-type asteroids in the 0.44 -- 0.65 $\mu$m range, while \citet{lantz2018} found a value of 0.05 \%/100 nm/deg in the 0.45 -- 2.45 $\mu$m range still on Near-Earth D-type asteroids. Even considering the largest spectral reddening coefficient for D-types, the spectral reddening effect should be lower than 1.2 \%/100 nm in the 5 -- 30° phase angle range where Gaia data are typically acquired.

We performed a first automatic classification. We automatically deleted points which were flagged 2, meaning not trustworthy, as well as the ones at the edges of the photometers (0.33, 0.64, and 1.05 $\mu$m) known to be affected by systematic problems. We applied a correction developed by \citet{Correction} of the artificial reddening, induced by solar analogs chosen to compute the relative reflectance, for the near-UV region. We visually inspected the spectra and some known problematic data, centered at 0.638$\mu$m, 0.418$\mu$m, and 0.990 $\mu$m \citep{galluccio:hal-04225764} were eventually excluded from the analysis if anomalous compared to the surrounding data. \\

To identify the taxonomic class, we compared the Gaia spectrophotometric data of a given asteroid with the mean spectra of the classes in the two aforementioned taxonomies. We perform a first automatic classification computing the $\chi ^2$ for each class, after interpolation of the Gaia spectra, as: $$\chi^2=\sum_{\lambda=0.314}^{\lambda=0.99}{\frac{{(R_{spectra}}_\lambda-{R_{class}}_\lambda)^2}{{\sigma_{class}}_\lambda^2}}$$

Where $R_{spectra}$ is the reflectance of the interpolated spectrum, $R_{class}$ is the reflectance of the mean spectrum for a given class, and $\sigma_{class}$ is the uncertainty for each point of the mean spectrum. What we would call the "best fit" are the classes for which the $\chi^2$ is the smallest. \\
We also computed the spectral slope between 0.550 and 0.814 $\mu$m, a range commonly used in the literature for visible spectral slope estimation, by linear regression of the data. We also computed the uncertainty of the slope as the standard deviation of the linear fit, meaning the square root of the diagonal terms of the covariance matrix. \\
This methodology produced  a list of potential D- and Z-types. From this list, we kept objects classified as primordial in both taxonomies, that is X-P, D-D or D-Z  in Bus-Demeo and Mahlke taxonomies.

Since D-types are known to be red, we considered only asteroids with a slope above 7 \%/1000~\AA\ for the D-type classification. As for the albedo, we used the constraints given by the definition of D-types according to \citet{Mahlke_2022}, meaning less than 12\%.

Similarly to the D-types, we imposed constraints on the slope and the albedo to obtain a list of potential P-types, and looked for those that were classified X- for Bus-DeMeo or P- for Mahlke. To avoid ambiguity with C-types, we kept objects with a slope above 2 \%/1000~\AA~as it was previously determined by \citet{fornasier2011}, and with an albedo between 4 and 7\%, as defined in Mahlke taxonomy. 

From those potential lists, we individually visually inspected each object to give a final classification and remove potentially known problematic points. 

For the study of the Hilda and Cybele populations, we did not do an initial automatic detection. Instead, we got the list of asteroids present in the populations from the \hyperlink{https://mp3c.oca.eu/}{MP3C database}, and directly classified the spectra present in Gaia for those populations. Nonetheless, we still applied the same criteria used for the main belt asteroids to the SNR. For Mahlke's classification scheme, X-type is only defined for objects for which the albedo is not known \citep{Mahlke_2022}. Hence, the reported X-types for the Cybele and Hilda in Mahlke taxonomy  is referred to bodies for which the albedo value in not available.

We included spectral slopes from the literature to complete the dataset of D-types in the main belt. We added 11 asteroids from \citet{oriel_2024_red} for the main belt. If for a given asteroid, both the spectral data were available in the Gaia catalogue and in the literature, we prioritized the data with the higher SNR, or, when comparable in SNR, we averaged the associated spectral slope values.

\subsection{Correlations}

To investigate correlations between different parameters, we have used Pearson's correlation coefficient. It varies between -1 and 1. If it is between 0.7 and 1, it means that it is strongly correlated; 0.5--0.7 indicates a moderate correlation. A faint correlation is for values between 0.2 and 0.5; anything under that means that there is no correlation. Conversely, if the value is negative, it means an anti-correlation (with the same ranges, in absolute values,  described previously). We computed this coefficient among surface properties (albedo and spectral slope), the diameter, and orbital parameters (semi-major axis, eccentricity, inclination, and aphelion). To determine the significance of the correlations computed, we computed their p-value. The closer the p-value is to zero, the more likely the correlation is significant.

\section{Main Belt}

\subsection{D- and Z-types}
\begin{figure*}
    \centering
    \includegraphics[width=0.9\textwidth]{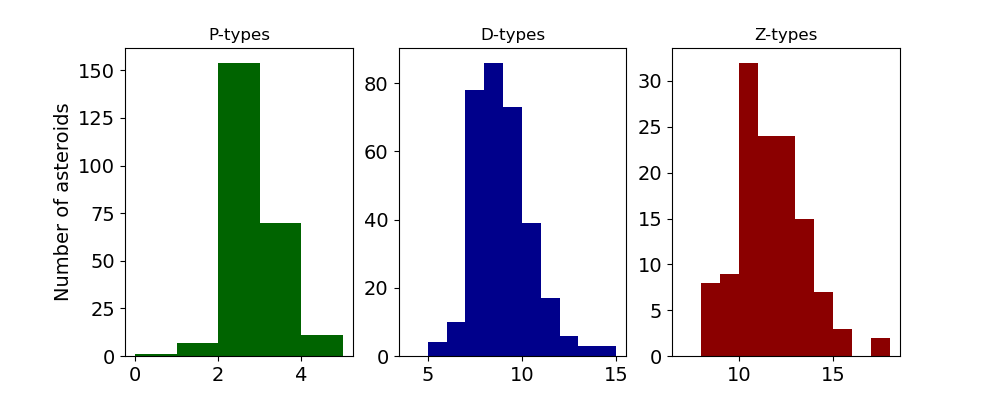}
    \caption{Histogram of the spectral slope distribution for (from left to right) P-, D-, and Z-types in the main belt. The bin size is of 1 $\%/1000$~\AA.}
    \label{fig:hist_slopes_mba}
\end{figure*}

From our analysis, we have found in the main belt 320 (318 uniquely from Gaia) D-types (0.88\% of Gaia catalog with SNR $>$ 20)  and 124 Z-types (0.34\%). For the distribution, we found 65 D-types in the inner, 88 in the middle, and 168 in the outer main belt, and for Z-types: 14 in the inner, 33 in the middle, and 77 in the outer main belt.  The limits of these regions are defined as 2.0 to 2.5 AU for the inner, 2.5 to 2.82 AU for the middle, and 2.82 to 3.2 AU for the outer main belt. \\

\begin{figure}
    \centering
    \includegraphics[width=0.9\hsize]{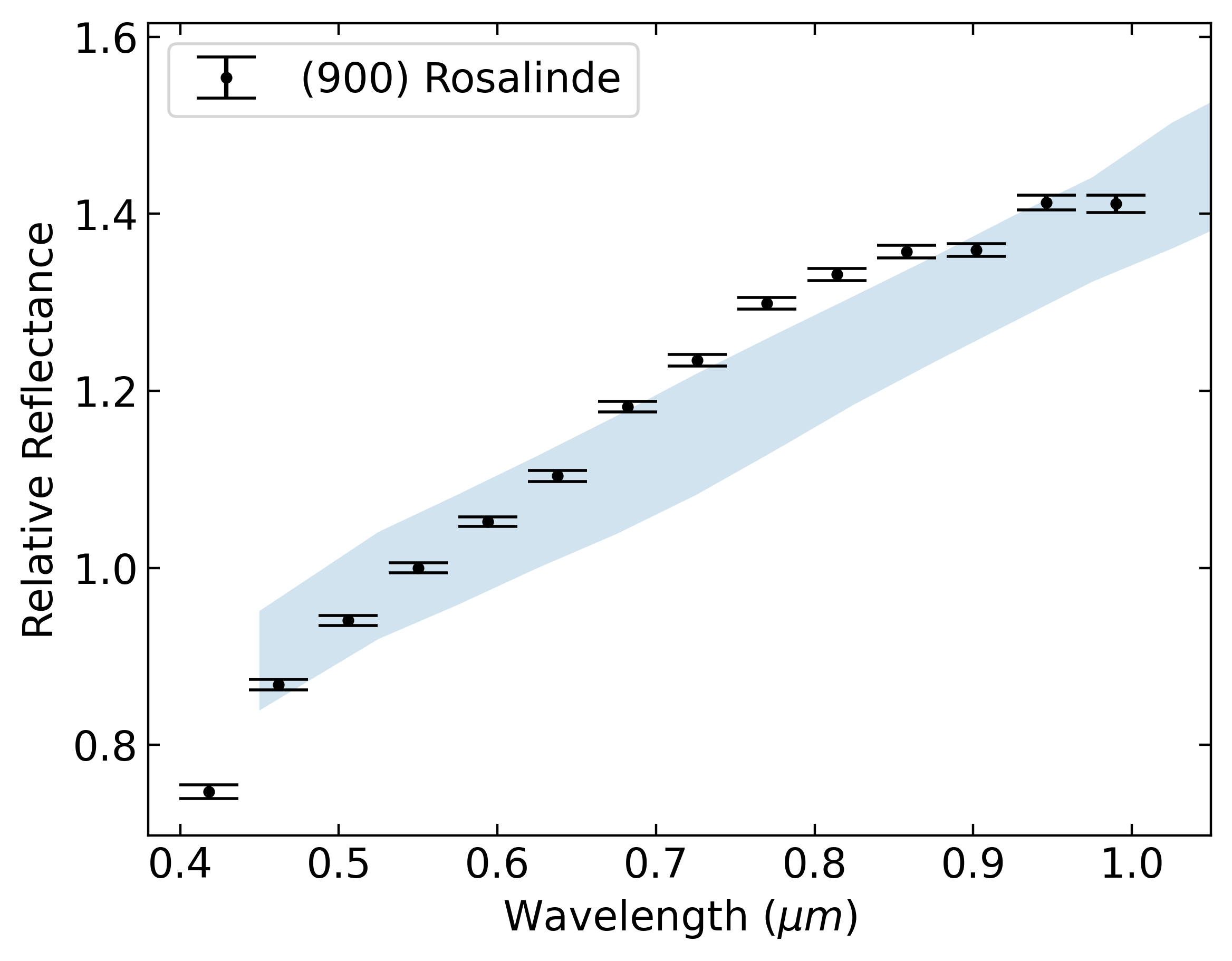}
    \caption{Spectrum of (900) Rosalinde from the Gaia spectral catalog. The filled blue region represents the mean spectrum of Mahlke's D-type.}
    \label{fig:ast900}
\end{figure}

The average slope is of $9.03 \pm 0.09$ \%/1000~\AA\ for D-types, and $11.68 \pm 0.16$ \%/1000~\AA\ for Z-types, respectively. Their slope distributions are shown in Fig. \ref{fig:hist_slopes_mba}. D-types peak at 8 \%/1000~\AA\ while Z-types peak at 11 \%/1000~\AA. D-types have a median slope of 8.76 and Z-types of 11.50  \%/1000~\AA. The reddest object of the dataset is (269) Justitia with a spectral slope of $17.70 \pm 0.56$ \%/1000~\AA. We confirm the peculiar red spectrum of this asteroid, already noticed by \citet{ast269}, who defined it as a TNO-like body and suggested a transneptunian origin.\\
Some D-types are as red as Z-types. This is because their spectra are steep in the 0.55–0.81 µm range, but flatten at longer wavelengths. An example of this spectral behavior is reported in Fig. ~\ref{fig:ast900}, concerning the spectrophotometry of (900) Rosalinde. We have identified 24 asteroids exhibiting this type of spectral behavior. Ten of these were observed for the first time by Gaia, three were reported by the MOVIS survey \citep{popescu2018a}, six were identified by the Sloan Digital Sky Survey (SDSS) \citep{SDSS}, three were detected by the SMASS-2 survey \citep{SMASS2}, one was identified by  the $S^3OS^2$ survey \citep{s3os2}, and one was identified by \citet{fieber2015}. Overall, their taxonomic classification derived from photometric surveys is often ambiguous, ranging from D-, S-, and A-types. Similar spectral behavior, that is a flattening after $\sim$ 0.8 $\mu$m, was observed on blue cold classical TNOs, and associated with a surface composition including refractory organics or caused by porosity and grain size effects \citep{seccull2021}. For TNOs, the flattening of otherwise very red spectra in the visible range appears more frequently at longer wavelengths, i.e., beyond 1-1.2 $\mu$m, as observed for the organic double-dip and cliff type TNOs, as defined by \cite{PinillaAlonso2024}.  \\

\begin{table}
    \centering
    \caption{D-types asteroids showing a less steep spectral slope beyond 0.8 $\mu$m.}
    \begin{tabular}{c c c c}
    \hline
    (567) Eleutheria & (673) Edda & (865) Zubaida & (900) Rosalinde\\
    (1612) Hirose & (1849) Kresak & (2105) Gudy & (4378) Voigt\\
    (6144) Kondojiro & (6212) Frantzthaler & (6336) Dodom & (7773) Kyokuchikenm\\
    (13504) 1998 RV12 & (15815) 1994 PY18 & (22106) Tomokoarai & (29503) 1997 WQ38\\
    (33776) 1999 RB158 & (40196) 1998 RM180 & (40730) 1999 SY12 & (48645) 1995 UF8 \\
    (51822) 2001 OB25 & (55956) 1998 HO100 & (76750) 2000 JX37 & (93832) 2000 WU77\\
    \hline
    \end{tabular}
    \label{tab:flat_d_type}
\end{table}

The mean albedo of D- and Z-types is practically the same, 7.9\%, with uncertainties of 0.1\% and 0.2\%, respectively. For the three regions in the main belt, we found:  9.1\% for D-types, and 9.6\% for Z-types in the IMB, 8.5 and 8.7\% in the middle main belt (MMB), and 7.2 and 7.6\% in the outer main belt (OMB). Therefore, inner belt D- and Z-types have higher albedos than in the rest of the main belt, as already noticed by  \citet{DeMeoDtype}, who reported an average albedo of 9.0\% for inner D-type asteroids. 
\begin{table}
    \caption{Number of D- and Z-types in each family.}
    \centering
    \begin{tabular}{c c c}
    \hline
         Family & D-types & Z-types\\
         \hline
         Eos &8 & 5\\
         Ursula & 5 & 4\\
         Eunomia & 9& 0\\
         Charis &4 &1\\
         Agnia & 4 &1\\
         Gersuind &4 &1\\
         Hygiea & 4&1\\
         Watsonia & 2 & 1\\
         Brangane & 2 &1\\
         Meliboa & 2 &1\\
         Tirela & 1 &1\\
         Emma & 0 &1\\
         Vesta & 0& 1\\
         \hline
    \end{tabular}

    \label{tab:families}
\end{table}

There are 45 D-types, and 19 Z-types that are members of 13 families (from \citet{Nesvorny2015}), including 13 members of Eos (8 D-, and 5 Z-types), 9 members of Ursula (5 D-, and 4 Z-types), 9 D-type members of Eunomia, 4 D-types and a Z-type from both the Gersuind and Hygiea families, 4 members of Watsonia (3 D-, and 1 Z-type), and 3 from Brangane and Meliboea families (2 D- and one Z-type each). There is also 4 D-types, one Z-type, and one D-type from Charis, Agnia. There is a D-type and a Z-type in Tirela. There is a Z-type from Emma and Vesta families (Table \ref{tab:families}).\\
Eos is primarily a K-type dominated family, Eunomia is mostly S-type,  Hygiea is a C-type one, Ursula is mostly composed of C- and X-types \citep{morate2018}, and Vesta is dominated by V-type asteroids \citep{Cellino2002,erasmus2019}. The Gersuind family is a more complex case. \citet{carruba2010} found a D-type (that is present in Gaia spectral catalog, (1609) Brenda, but was discarded due to its albedo being higher than 0.12), yet they believe that it is an interloper. Tirela was first believed to be a D-type family based on its parent body (1400) Tirela's spectral properties \citep{lazzaro2004}, but further observations tend more toward an L-type dominated family \citep{mothe2008}. In conclusion, in the indicated families, D-types are very likely interlopers. 
\begin{table}
    \caption{Pearson correlation coefficient and p-value for D- and Z-types.}
    \centering
    \begin{tabular}{c c}
    \hline
       Variables  & Correlation coefficient ($P_r$)  \\
       \hline 
       \multicolumn{2}{c}{D-types}\\
       \hline
        $a$ vs $p_v$ & -0.36 (1.74$\times 10^{-11} $)\\ 
        $a$ vs $e$ & -0.28 (3.4 $\times 10^{-7})$\\
        $a$ vs $D$ & 0.2 (0.00027)\\
        $D$ vs $p_v$ & -0.23 (3.9$\times 10^{-5}$)\\
        \hline
        \multicolumn{2}{c}{Z-types}\\
        \hline
        $a$ vs $p_v$ & -0.34 (0.00012)\\
        $a$ vs $D$ & 0.29 (0.0013)\\
        $D$ vs $p_v$ & -0.38 (1.7 $\times 10^{-5}$)\\
        \hline
    \end{tabular}
    \label{tab:corrD}
\end{table}

\begin{figure}
    \centering
    \includegraphics[width=0.5\textwidth]{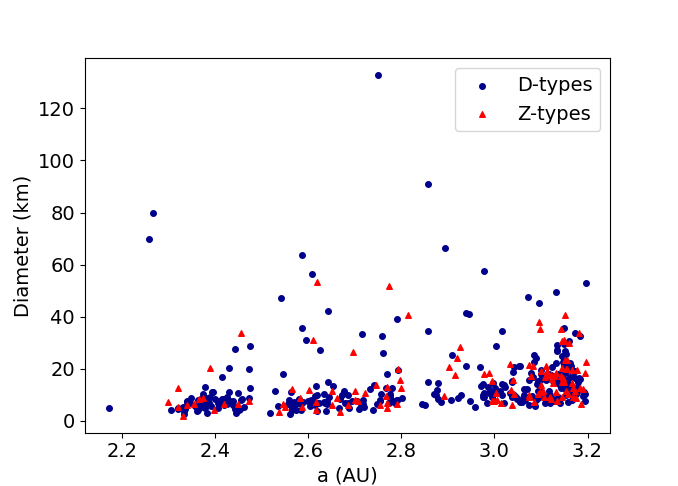}
    \caption{Diameter as a function of the semi-major axis. The D-types are represented by blue dots, while Z-types are shown with red triangles.}
    \label{fig:diam_a_dz_mba}
\end{figure}

We look for correlations among physical parameters and orbital elements. The computed Pearson correlation coefficients for D- and Z-types are reported in Table \ref{tab:corrD}. Some of them can be partially attributed to observational bias. The weak correlation (for both D and Z) between the diameter and the semi-major axis ($r$ = 0.20, $P_r$ = 0.00027), is likely an observational bias, because it is difficult to observe smaller objects when they are far from the Sun. However, if present, asteroids larger than 30-40 km should be observed, whereas there is an evident lack of them in the inner belt (Fig. \ref{fig:diam_a_dz_mba}), with the notable exception of (72) Feronia, an 80 km sized D-type asteroid \citep{Jules2023}.\\
The lack of large D-type asteroids in the inner main belt was already noticed by \citet{DeMeoDtype}, who proposed that the observed D-types were implanted there by the Yarkovsky effect, and hence only the smaller ones passed the 3:1 mean motion resonance. They proposed as an alternative explanation that the D-type parent body experienced a collision near the resonance, giving it enough energy to pass through the resonance into the inner main belt. \citet{volkrou2016} suggested thermal destruction on the path to the inner main belt, causing the escape of volatiles, hence resulting in smaller objects as they get closer to the Sun. This correlation was noted as well by \citet{Gartrelle2021b}. Nevertheless, it was a stronger correlation for non-Trojan D-types than ours. \\
The biggest object of our dataset is (308) Polyxo, with $D=132.7 \pm 1.0$ km. It is traditionally classified as a T-type \citep{bus2002,demeo_tax, Belskaya2017, Kwon2022}. Yet, Gaia's spectrum, which has a high SNR of 908, does not fit in the slope's range of T-types ($S$=7.549 $\pm$ 0.064 \%/1000~\AA), nor does it show the concaving down around 0.75 $\mu$m descriptive of T-types \citep{Bus99,demeo_tax}. We classify it as D-type on both taxonomies (of note of evidence, the T-class does not exist in Mahlke taxonomy). The Gaia redden spectrophotometry compared to literature data may be partially attributed to instrumental effects, which however mostly impact the wavelengths higher than 950 nm. The Polyxo Gaia data may be redder compared to literature data also because of potential spectral phase reddening effects, which however we cannot quantify because Gaia data are averaged over different epochs.  Finally, we cannot exclude surface heterogeneity across Polyxo surface, which may eventually be investigated with devoted observations covering different rotational phases. \\
The Gaia reddening effect has been reported mostly for A-, S-, K-, and L-types, independently of physical parameters or magnitude, and does not seem to impact as much featureless objects, such as C- and P-types \citep{galinier:tel-04814227}. Some of the asteroids that we classified as D-types may potentially be L-types misclassified because of the Gaia spectral reddening reported in the literature \citep{galluccio:hal-04225764}. However we applied a limit on the albedo of less than 12\% to filter them. Nonetheless, for cases where the albedos have large uncertainties, it is possible to have some ambiguities, and misclassifications. For instance, we classified the asteroid (611) as D-type using Gaia data. However, different classifications have been reported in the literature: M-type using VNIR spectra by \citet{Mahlke_2022}; L-type by \citet{bus2002} and \citet{devogele2018}; and S-type by \citet{tholen1989}.

We also found a weak anti-correlation, both for D- and Z-types between the semi-major axis and the albedo, with low significance values. It cannot be caused by an observational bias, since we tend to detect brighter asteroids when they are far from the Sun. It differs from the finding of \citet{Gartrelle2021b}, which found an increase of the albedo with the semi-major axis for non-Trojans D-types. This anti-correlation was also noticed by \citet{DeMeoDtype}, and attributed to rejuvenating processes caused by space weathering. \\ 
We found a faint anti-correlation, for both spectral types, between the diameter and the albedo, meaning the smaller the asteroids are, the brighter. It seems to be an observational bias, as smaller asteroids are easier to detect and observe when bright. It could also mean that smaller bodies have a fresher hence brighter material at their surface due to collisions/resurfacing processes. \\
No correlations were found regarding the slope in the visible, similarly to the findings of \citet{Gartrelle2021b}. The only exception is for Z-type members of families, but, as previously shown, likely interlopers. These bodies show a  moderate correlation between the eccentricity and the slope, as well as a weak one with the inclination. Yet, considering there are only 20 objects in this category, the statistics are very poor. Moreover, we do not report any correlations between the inclination and the semi-major axis in the primordial main belt asteroids, contrary to \citet{Gartrelle2021b} findings. However, in their study,  \citet{Gartrelle2021b} includes Jupiter Trojans, which have inclinations up to 40 $^{\circ}$, while we consider only main belt D-type, which have lower inclinations.

\subsection{P-types}

\begin{figure}
    \centering
    \includegraphics[width=\linewidth]{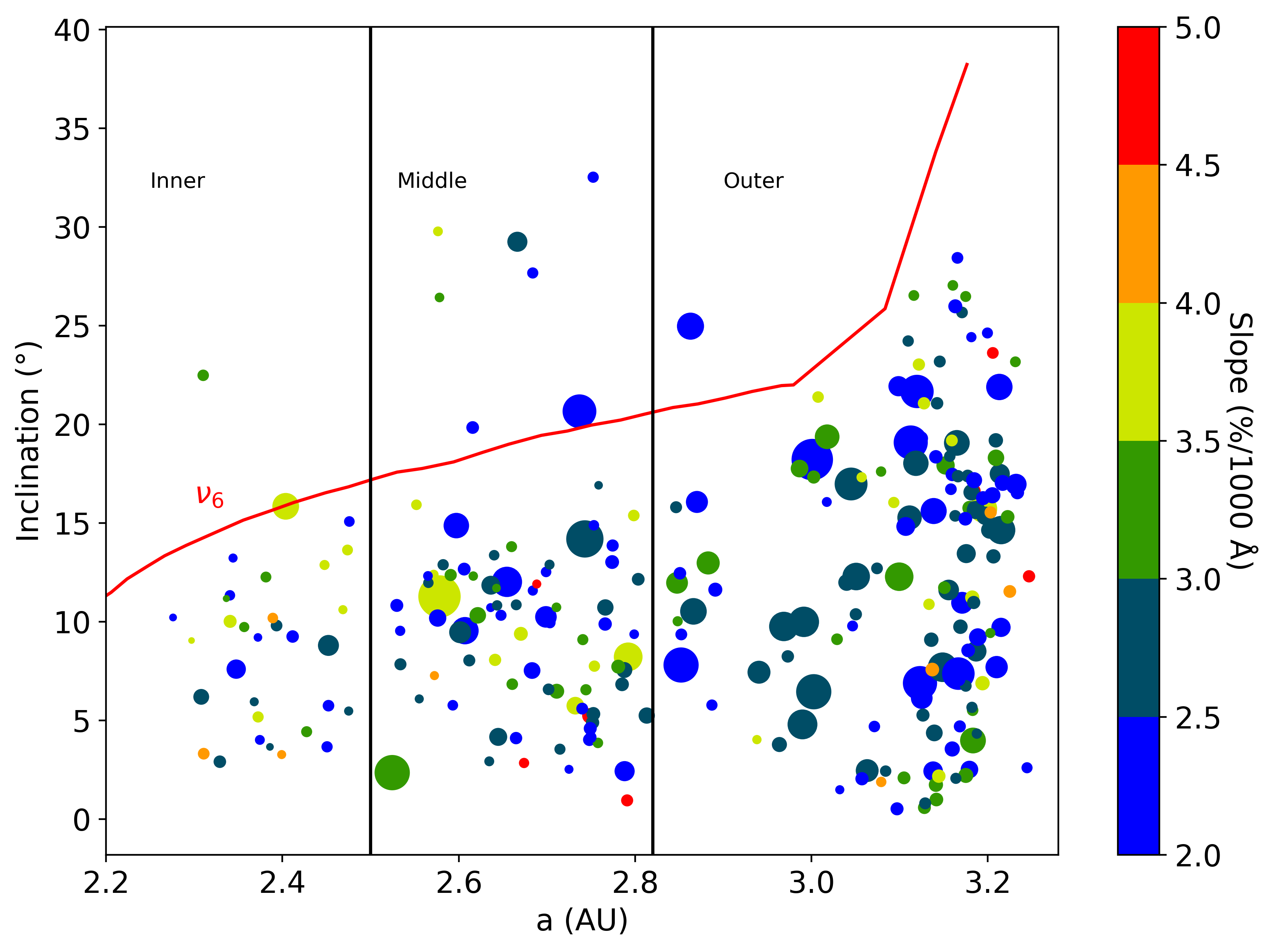}
    \caption{Inclination as a function of the semi-major axis for P-types in the main belt. Symbol size is proportional to the diameter. The colorbar represents the spectral slope. The red line is the $\nu_6$ secular resonance position in this space \citep{morbidelli2010}.}
    \label{fig:pos_p}
\end{figure}

We have classified 243 P-types in the main belt. 32 of them are in the IMB (12.07\% of all P-types in the dataset), 87 in the MMB (32.83\%), and 124 in the OMB (46.79\%). In a previous study, \citet{Mahlke_2022} classified 73 P-types in the main belt, with similar percentages across different regions: 11 in the IMB (15.07\% of P-types in the main belt), 26 in the MMB (35.62\%), and 36 in the OMB (49.32\%). 
The P-type average slope value from our dataset is  $2.83 \pm 0.04 \%/1000$~\AA, and their spectral distribution peaks in the 2-3 $\%/1000$~\AA\ range (Fig. \ref{fig:hist_slopes_mba}). Their average  albedo is 5.48 $\pm$ 0.05\%.

The only correlation found is a weak one between the inclination and the semi-major axis ($r$=0.21, $P_r$=0.00069), as shown in Figure \ref{fig:pos_p}. Hence, the farther away they are from the Sun, the more inclined they tend to be. It seems to be related to the $\nu_6$ secular resonance with Saturn, which impacts asteroids eccentricity and causes gaps within the main belt \citep{morbidelli2010}. 
The inclination trend of P-types is related to the $\nu_6$ resonance position and its steep increase at $\sim$3 AU. This resonance would prevent objects from reaching higher inclinations throughout the main belt.\\
Despite the lack of correlation between the diameter and the heliocentric distance, Figure \ref{fig:pos_p} shows that there are more large objects in the outer main belt than in the inner one, similarly to D- and Z-types. Despite the fact that this is likely an observational bias, their mean diameter is $24.2 \pm 1.8$  km, that is larger than that of D/Z-types observed at similar distances (Table \ref{tab:SolSys}).

\begin{table}
    \caption{Number of P-types in each family.}
    \centering
    \begin{tabular}{c c c c}
    \hline
         Family & P-types & Family & P-types \\
         \hline
        Ursula&12 &Themis&9\\
        Euphrosyne&7 &Adeona &7\\
        Padua& 6 & Alauda &6\\
        Klio & 5 & Erigone&4\\
        Hygiea&3 &Flora & 2\\
        Brucato&2 &Chimaera &2\\
        Karma & 1 & Dora & 1\\
        Lixiaohua&1 &1993FY12 &1\\
        Phocaeo&1 & Nysa-Polana&1\\
        Mitidika &1 & 19998YB3&1\\
        Terpsichore&1 &Astrid &1\\
        Yakolev &1 &Marconia &1\\
        Emma & 1 &Ino & 1\\
        Sulamitis &1 & Misa & 1\\
        Armenia &1 &Meliboea &1\\
        Rafita &1 &Ino&1\\
        
         \hline
    \end{tabular}

    \label{tab:familiesp}
\end{table}

There are 85 family members within the MBA P-type dataset, as shown in Table \ref{tab:familiesp}: 12 members of Ursula, 9 of Themis,7 of Euphrosyne,7 of Adeona, 6 members of Padua and Alauda, 5 of Klio, 4 of Erigone, 3 from Hygiea, 2 of Flora, Brucato, and Chimaera, and one member for 20 other different families.\\
Themis is mostly composed of C- and B-asteroids \citep{mothe2005, kaluna2016, florczak1999, deleon2012, fornasier2016}, Adeona of C-types \citep{mothe2005}, Padua of X- and C-types \citep{mothe2005}. Euphrosyne consists of C-types, yet 1.2\% of Euphrosyne members have been classified as D-types \citep{carruba2014}. Erigone is a primordial family composed of X- and C-types \citep{morate2016}. The presence of P-types within primordial families such as Ursula, Klio, or Themis is not surprising, especially since X-types in Bus-DeMeo taxonomy cover P-types in Mahlke taxonomy. The presence of an important fraction (35\%) of the main belt P-types in C- and X-families hints at a common origin, with spectral variations among family members likely related to space weathering (SW).

\section{Cybele} \label{cyb}

\begin{figure}
    \centering
    \includegraphics[width=0.5\textwidth]{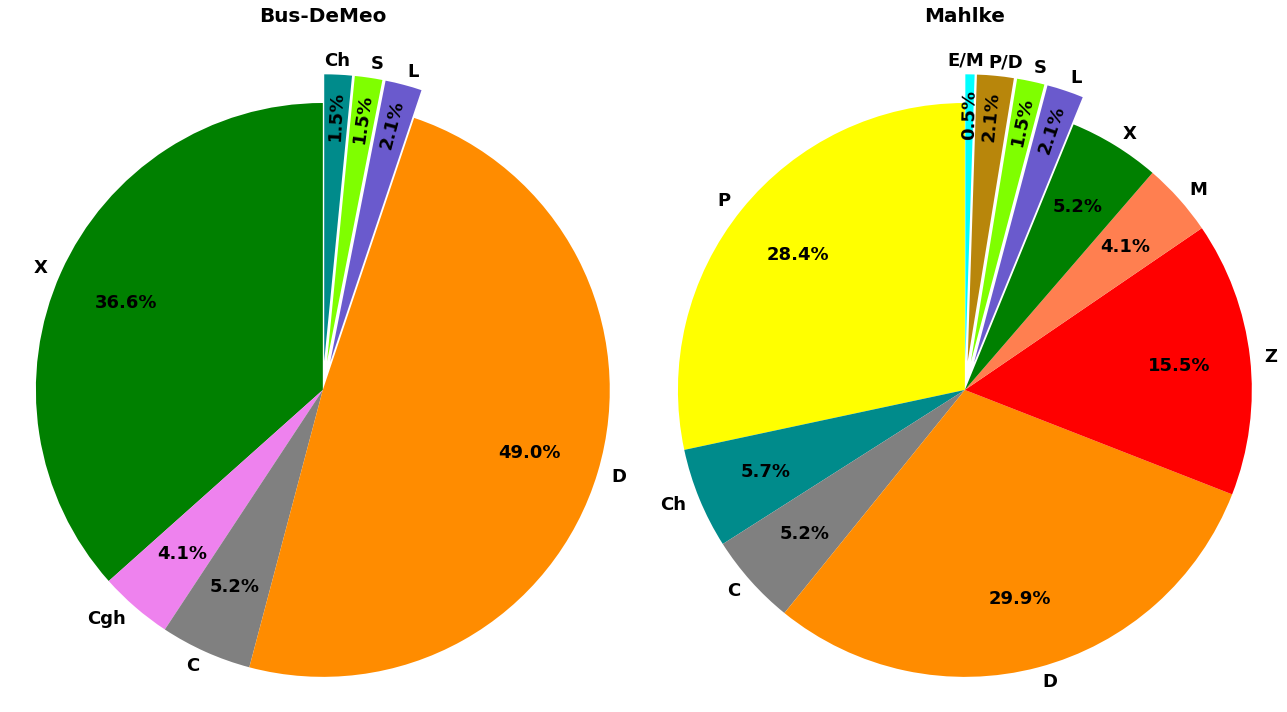}
    \caption{Pie charts presenting the taxonomic distribution for the Cybele population : Bus-DeMeo taxonomy on the left panel and Mahlke taxonomy on the right one.}
    \label{fig:pie_cyb}
\end{figure}

In the Gaia DR3 catalog, we identify 193 Cybele asteroids with SNR $>$ 20. Their taxonomic classification is as follows: in Bus-DeMeo
taxonomy, 49.0\% are D-types, 36.6\% X-types, 5.2\% C-types, 4.1\% Cgh-, 2.1\% L-types, and 1.5\% both of Ch and S-types; In Mahlke taxonomy, 29.9\% D-types, 28.4\% P-types, 15.5\% Z-types, 5.7\% Ch-types, 5.2\% C- and X-types, 4.1\% of M-types, 2.1\% of L- and P/D- types, 1.5\% S-types, and 0.5\% E/M- (Fig. \ref{fig:pie_cyb}).
\citet{Mahlke_2022} in their study reported  57.1\% of P-types, 23.8 of D-types, 9.5\% of C-types, and 9.5\% of Ch-types, but their sample was limited to 21 Cybele asteroids. We find more diversity, including the presence of Z-types, and most importantly, that the majority of the Cybele population consists of D- and P-types. In addition, we identify three S-type objects, unexpected at these heliocentric distances.

\begin{figure}
    \centering
    \includegraphics[width=0.5\textwidth]{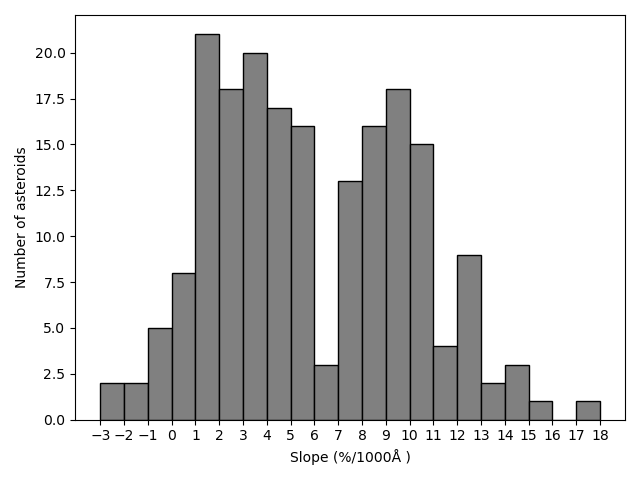}
    \caption{Spectral slope distribution of the Cybele population. The bin size is of 1 $\%/1000$~\AA.}
    \label{fig:hist_cyb}
\end{figure}

The mean spectral slope value of the Cybele population is $5.93 \pm 0.29$ \%/1000~\AA. Their average diameter is $41.1 \pm 3.2$ km,  and their mean albedo of $6.9 \pm 0.3$\%. The slope distribution is reported in Fig.\ref{fig:hist_cyb}.  It shows a bimodality, with two peaks located at 2-4 \%/1000~\AA\ and 8-10 \%/1000~\AA\, respectively, reflecting the presence of moderately sloped asteroids such as C- and P-type, and very red ones like D- and Z-type. This bimodality was already reported in the literature \citep{gilhutton2010, tinaut2024}. These last authors found a bimodal distribution for larger Cybele and the Hilda asteroids having absolute magnitude $H$ comprised between 8 and 12, but not for smaller  objects with  12 $< H < $ 16.

\begin{table}
    \caption{Pearson correlation coefficient and p-value for the Cybele population.}
    \centering
    \begin{tabular}{c c}
    \hline
     Variables  & Correlation coefficient ($P_r$)  \\
    \hline 
    $D$ vs $p_v$  & -0.28 (0.00021)\\
    $D$ vs $S$ & -0.35 (1.9$\times10^{-6}$)\\
    \hline
    \end{tabular}
    \label{tab:corrcyb}
\end{table}

\begin{figure*}
    \centering
    \includegraphics[width=0.9\textwidth]{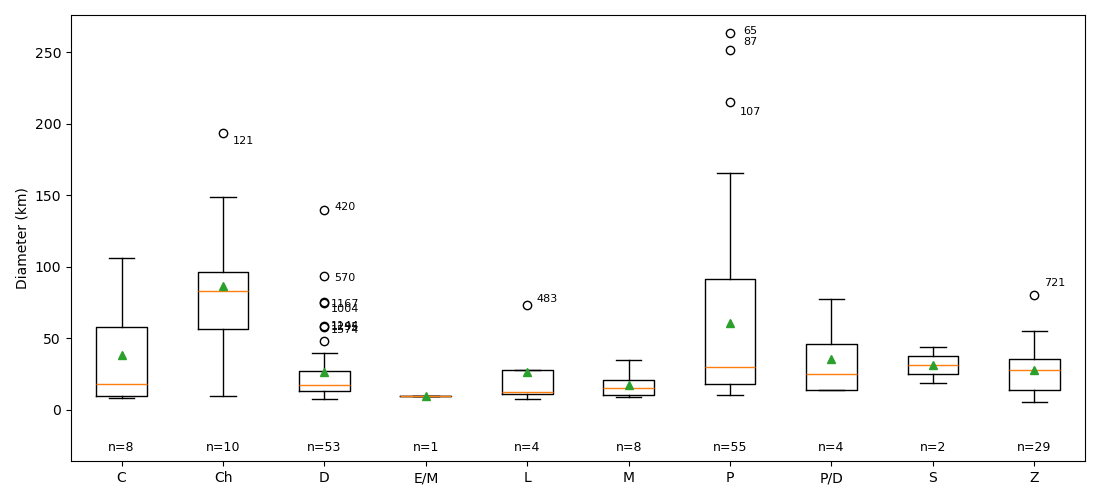}
    \caption{Boxplots showing the diameter for each Mahlke class within the Cybele population. Outliers are not included in the box and are shown by the dots. The green triangles represent the mean for each class, and the orange lines show the median. The boxes extend from the first to the third quartile.}
    \label{fig:boxplots_cyb}
\end{figure*}

The correlations found for the Cybele population are reported in Table \ref{tab:corrcyb}. There is a weak anti-correlation between the diameter and the albedo, as well as between the diameter and the spectral slope. Therefore, the larger the objects are, the darker and bluer they tend to be. Hence, redder objects (D-types) are smaller than less red (P-types) ones. The anti-correlation between the slope and the diameter was previously reported by \citet{lagerkvist2005} and  \citet{gilhutton2010}. One way to explain this would be that D-types are composed of more fragile material and more likely to experience collisions than P-types \citep{lagerkvist2005}. It could also mean that P-types were formed in situ, hence experienced less collisions, whereas D-types could have been implanted from the outer Solar System, experiencing more perturbations and collisions. \\
The weak anti-correlation between the diameter and the albedo means that the bigger they are, the darker they get. In Fig. \ref{fig:boxplots_cyb}, we report the diameter versus the taxonomic classes, indicating the mean, median, and quartiles for the diameter for each class identified in the Cybele population. The largest bodies belong to the P-class, which is also characterized by low albedo values, comprised between 4 and 7\% \citep{Mahlke_2022}.

There are two families in the Cybele region: Sylvia, with 255 known members, and Ulla, with 26 known members \citep{Nesvorny2015}. There are 14 members of the Sylvia family within our Cybele dataset, and their classification is as follows: 5 P-types (35.7\%), 4 D-types (28.6\%), 3 X-types (21.4\%), one C-type, and one L-type (7.1\% for each class). It is surprising to have D-types as the second most frequent taxonomy, as the Sylvia family is defined a C/X-type dominated family \citep{carruba2015}. Yet, it is important to note the small sample of the Sylvia family we have, hence is prone to statistical errors. For the Ulla family, we have in the Gaia catalogue only two members, both belonging to the P- class. 

\section{Hilda}

We finally analyze the Hilda population. We have a sample of 180 asteroids with SNR $>$ 20 in the Gaia DR3. 

\begin{figure}

    \centering

    \includegraphics[width=0.5\textwidth]{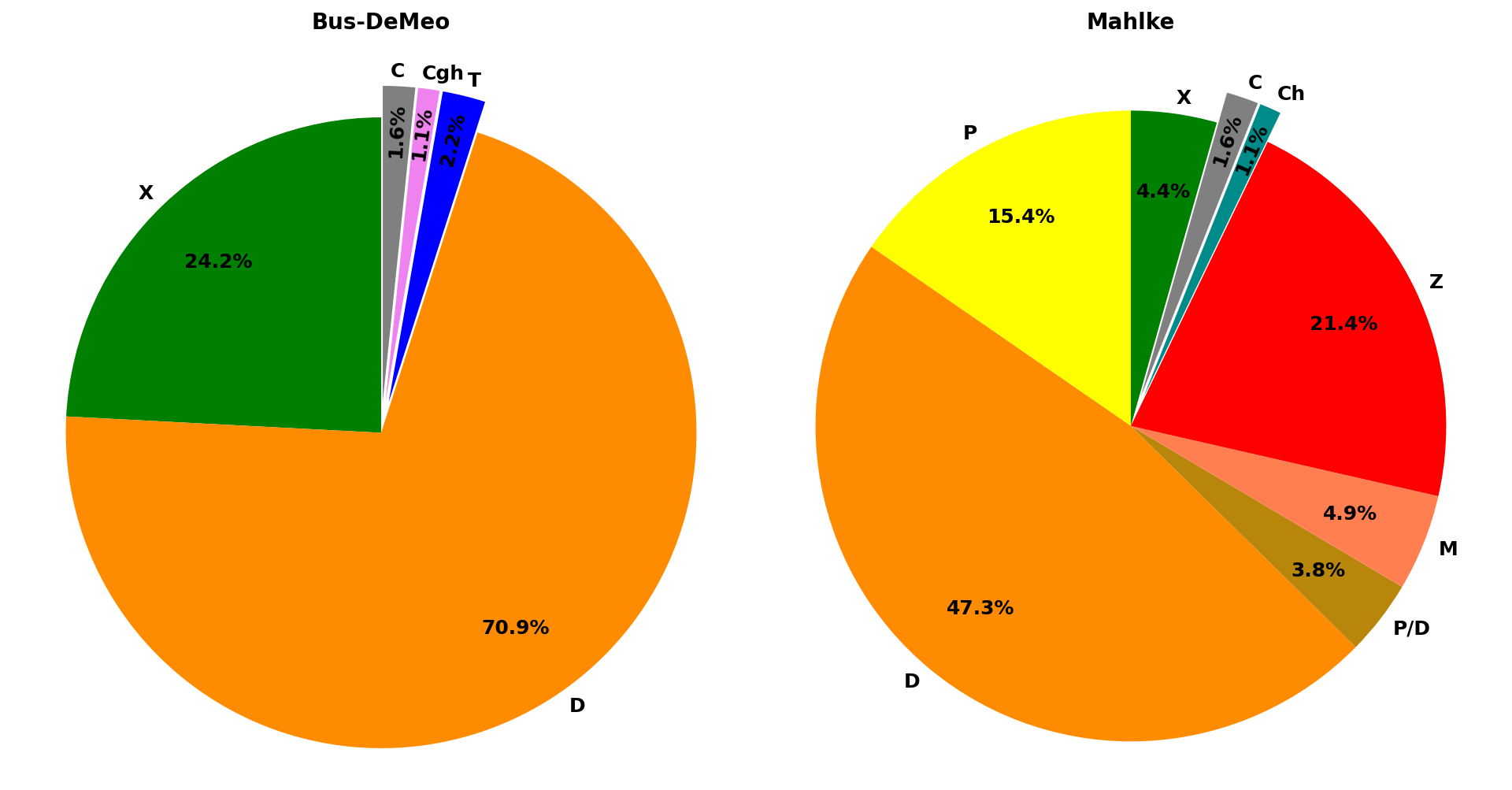}

    \caption{Pie charts showing the taxonomic distribution for the Hilda population, according to Bus-DeMeo taxonomy (left panel) and Mahlke (right panel) taxonomies.}

    \label{fig:pie_Hildas}

\end{figure}

Their taxonomic composition is shown in Fig. \ref{fig:pie_Hildas} for both Bus-DeMeo and Mahlke's taxonomies. In Bus-DeMeo, we classify 70.9\% of D-types, 24.2\% of X-types, 2.2\% of T-types, 1.6\% of C-types, and 1.1\% of Cgh-types, while in Mahlke's, we have 47.3\% of D-types, 21.4\% of Z-types, 15.4\% of P-types, 4.9\% of M-types, 4.4\% of X-types (for these bodies the albedo was not available), 3.8\% of P/D-types, 1.6\% of C-types, and 1.1\% of Ch-types. \\
Similarly to the Cybele, \citet{Mahlke_2022} reported in the Hilda 47.1\% of D-types, 35.5\% of P-types, 5.9\% of C-types, and 5.9\% of M-types, on a sample of 34 objects. We report more taxonomic diversity and, once again, the presence of Z-types.

The presence of M-types is surprising, even if they represent a small portion of the Hilda dataset (4.92\%), as they are associated to metallic cores and differentiated bodies \citep{gaffey79}, yet the Hilda are too far from the Sun to receive enough heat for differentiation. If we consider that the D- and Z-types originate from TNOs, and M-types from the main belt, it could hint at mutual mixing both from the inner and external parts of the Solar System. In Mahlke's classification scheme, E-, M-, and P-types are differentiated mostly albedo-wise, and through infrared features \citep{Mahlke_2022}. Considering the errors associated with the albedo values, and the fact that Gaia spectra cover only the visible range, M-classification has to be considered carefully and confirmed by additional observations. \\
Overall, the Hilda show less taxonomic diversity than the Cybele (Section \ref{cyb}), they  present a larger portion of primitive asteroids (D-, Z-, and P-types), and have fewer carbonaceous objects. \\

\begin{figure}

    \centering

    \includegraphics[width=0.5\textwidth]{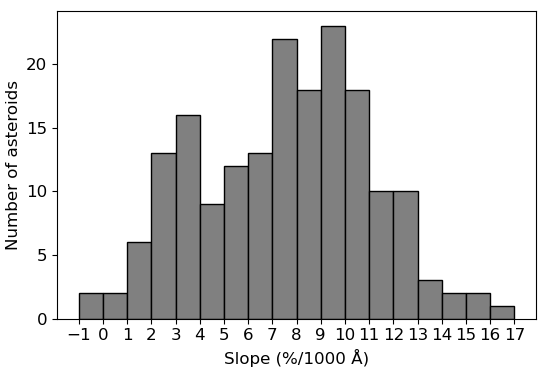}

    \caption{Spectral slope distribution for all the Hilda. The bin size is of 1 $\%/1000$~\AA. }

    \label{fig:hist_slope_Hildas}

\end{figure}

The mean spectral slope of the Hilda population is $7.49\pm0.26$ \%/1000~\AA. We report their slope distribution in Fig. \ref{fig:hist_slope_Hildas}. We find a slight bimodality, with two peaks located at 3 \%/1000~\AA, and in the 8--10\%/1000~\AA\ range. This bimodality was first observed in the analysis of SDSS data  by \citet{gilhutton2008}, and was more pronounced than in our analysis. It is consistent with the findings of \citet{wong2017_spectra}, which classified 26 Hilda spectra in the near-infrared into two categories: LR for less-red, and R for red. 

\begin{figure*}

    \centering

    \includegraphics[width=0.9\textwidth]{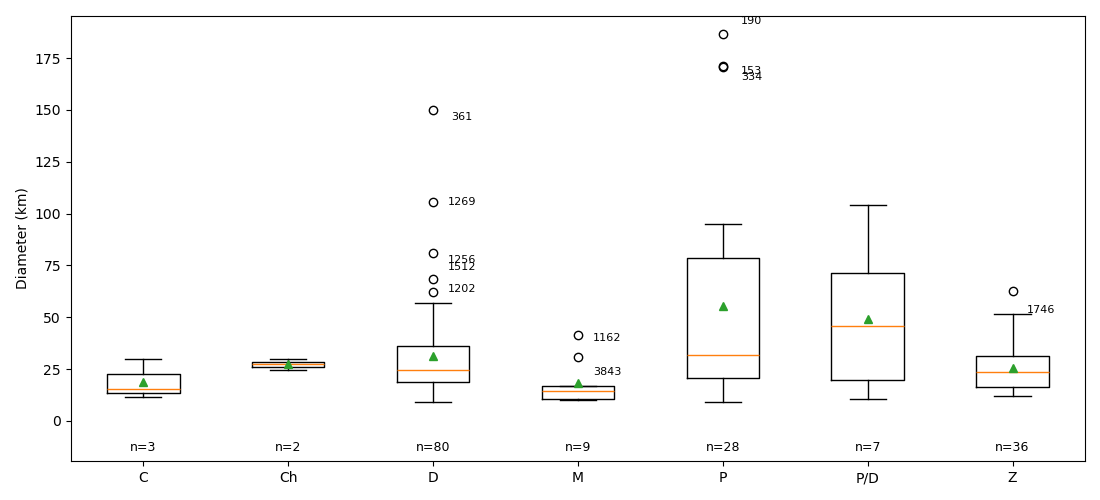}

    \caption{Boxplots of the diameter for each Mahlke's class for the Hilda population. The outliers are not taken into account in the box and are shown by the dots. The green triangles represent the mean for each class, while the orange lines show the median. The boxes go from the first to the third quartile.}

    \label{fig:boxplots_hil}

\end{figure*}

The distribution, mean, median, and quartiles of the Hilda's diameter for each Mahlke's class is presented in Fig. \ref{fig:boxplots_hil}. Similar to Cybele, P-types show a large range of sizes and contain the biggest asteroids.

\begin{table}

    \caption{Pearson correlation coefficient and p-value for the Hilda population}

    \centering

    \begin{tabular}{c c}

    \hline

     Variables  & Correlation coefficient ($P_r$)  \\

    \hline 

    $a$ vs $e$ & 0.32 (1.4 $\times 10^{-4}$)\\

    $D$ vs $p_v$  & -0.27 (0.00053)\\

    $D$ vs $S$ & -0.26 (0.00076)\\

    \hline

    \end{tabular}

    \label{tab:corrHildas}

\end{table}

\begin{figure}

    \centering

    \includegraphics[width=0.5\textwidth]{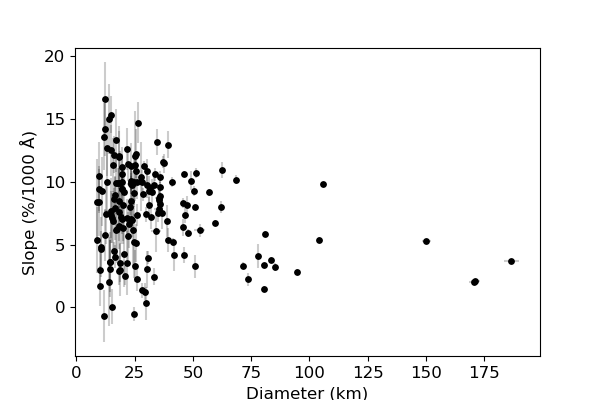}

    \caption{Spectral slope vs the diameter for the Hilda.}

    \label{fig:slope_diam_Hildas}

\end{figure}

Table \ref{tab:corrHildas} reports the significant Pearson correlation and p-value coefficients. There is a weak but significant (p-value = 1.4 $\times10^{-4}$) correlation between the semi-major axis and the eccentricity, meaning the farther from the Sun, the more eccentric Hilda asteroids are.  This correlation is driven by the 7 objects located between 3.85 and 3.91 AU (Fig.~\ref{fig:tiss_pos}) which have low inclination ($i$ < 6$^{\circ}$) and eccentricity ($e$ < 0.13). They do not share the same taxonomic types (2 P-types, 3 D-types, 1 Z-type, and 1 X-type), and they have a large range of diameters (going from 31 km up to 171 km). Their only similarities are their orbital properties. If we remove those seven objects, the correlation drops to 0.13, with a p-value of 0.092. Two of them, (334) Chicago and (8712) Amos, are reported to exhibit chaotic behavior \citep{ferraz92}. We could suspect that they are overall unstable, despite their low eccentricity and inclination. \\
The diameter is weakly anti-correlated with the albedo and the spectral slope, similarly to the Cybele population, even if the correlation is slightly less strong for the Hilda. For both those relationships, the p-value is low. Hence, larger Hilda would tend to be darker and bluer. It is directly caused by the fact that similarly to Cybele's, larger Hilda asteroids are P-types, whereas D-types are smaller. Figure \ref{fig:slope_diam_Hildas} illustrates the anti-correlation between the slope and the diameter, which was previously reported in the literature \citep{dahlgren1995,dahlgren1997,DePra2018}.  This distribution is consistent with the diameter distribution among Hilda asteroids (Fig. \ref{fig:boxplots_hil}). P-types include the larger asteroids, while D-types are on average smaller in size but also include a few large bodies (361 and 1269). Similarly to Cybele, \citet{dahlgren1997} suggests that D-types have a more fragile composition leading to more collisional fragments than P-types. Another hypothesis these authors discuss is variations related to heating mechanisms, such as electrical induction heating from the solar wind in the earlier phases of the Solar System. This process acts mostly on large objects. \citet{dahlgren1997} suggest that parent bodies in the Hilda asteroids were heated from the core, and then experienced collisions. In this hypothesis, the P-types would be the exposed interiors that went through more heating, whereas the D-types would be the outer layers.  

The Hilda region hosts two dynamical families: the Hilda family and the Schubart's one \citep{Nesvorny2015}. Within our dataset, we have 35 members of the Hilda family, and 15 of the Schubart family. 

\begin{figure}

    \centering

    \begin{subfigure}{0.5\textwidth}
       \caption{Hilda}
        \includegraphics[width=\textwidth]{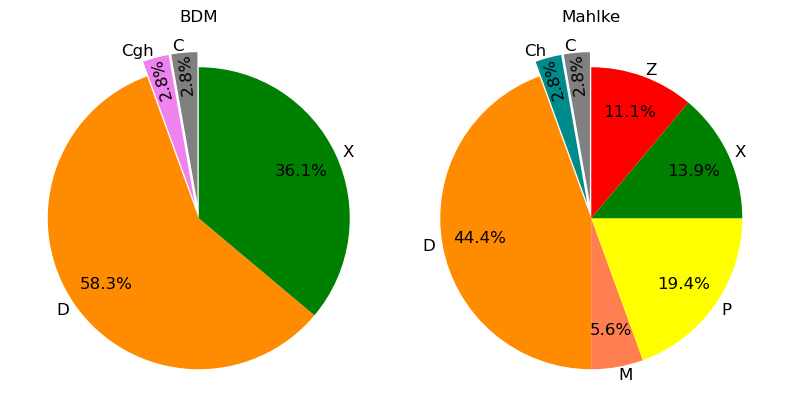}

    \end{subfigure}

    \begin{subfigure}{0.5\textwidth}
        \caption{Schubart}
        \includegraphics[width=\textwidth]{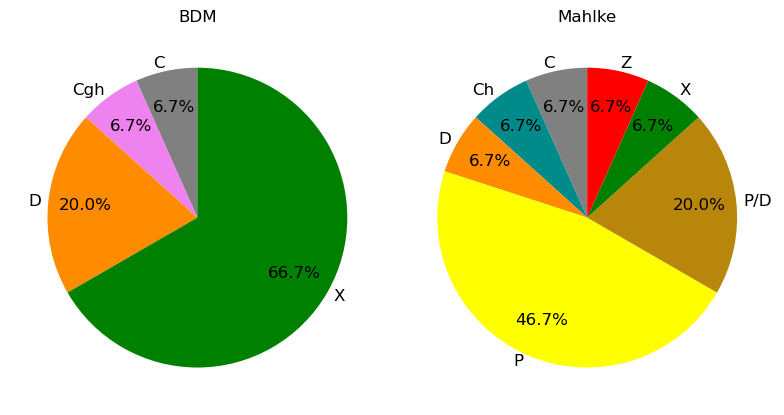}

    \end{subfigure}

    \caption{Pie charts showing the BDM (on the left) and Mahlke (on the right) taxonomies for Hilda (panel (a)) and Schubart (panel (b)) families.}

    \label{fig:pies_fam}

\end{figure}

Their classification is shown in Fig. \ref{fig:pies_fam}. For the Hilda family, in BDM, there are 58.3\% of D-types, 36.1\% of X-types, and 2.8\% of Cgh- and C-types. In the Mahlke scheme, there are 44.4\% of D-types, 19.4\% of P-types, 16.9\% of X-types, 11.1\% of Z-types, and 2.8\% of Ch and C-types. For the Schubart family in the BDM classification, it is composed of 66.7\% X-types, 20.0\% D-types, and 6.7\% Cgh- and C-types. In Mahlke, it is 46.7\% P-types, 20.0\% P/D-types, and 6.7\% of X-, Z-, C-, Ch-, and D-types. The Hilda family is dominated by D-types , whereas the Schubarts' one by P-types. The Schubart family contains more C-complex, consistent with what was reported in previous studies \citep{grav2012}. 

The mean albedo for the Hilda family is $6.3\pm 0.3$ and for the Schubart one $4.2 \pm 0.2$\%.

\section{P-, D-, Z-types in the Solar System}

We study the distribution of primordial asteroids, namely P-, D-, and Z-types, following Mahlke's classification scheme, in the Solar System. To complete our analysis on primordial asteroids, we also include the results on a sample of 519 Jupiter Trojans based on Gaia DR3 analysis and literature data published in \citet{Fornasier2025}. The authors have classified Jupiter Trojans present in Gaia's catalog both in BDM and Mahlke's taxonomy, finding more spectral heterogeneity in L4, but caused by family members, in particular by the Eurybates family ones. Both swarms are dominated by red and organic-rich primordial asteroids (72\% in L4 and 88\% in L5), with an important fraction of Z-types (32\% in L4 and 46\% in L5). The spectral distribution shows no bimodality, contrary to what is found in the literature on smaller samples \citep{Szabo_2007, Emery_2024}. On the basis of these results, \citet{Fornasier2025} suggested that Trojans originated from a common source in the outer Solar System, with spectral variability associated with evolutionary processes like collisions and space weathering. They found analogies among Trojans and TNOs, suggesting that the Trojans are captured TNOs, and notably that Centaurs and scattered TNOs are close to the Trojans' properties in terms of visible spectral slope and average albedo value.

\begin{figure*}

    \includegraphics[width=\textwidth]{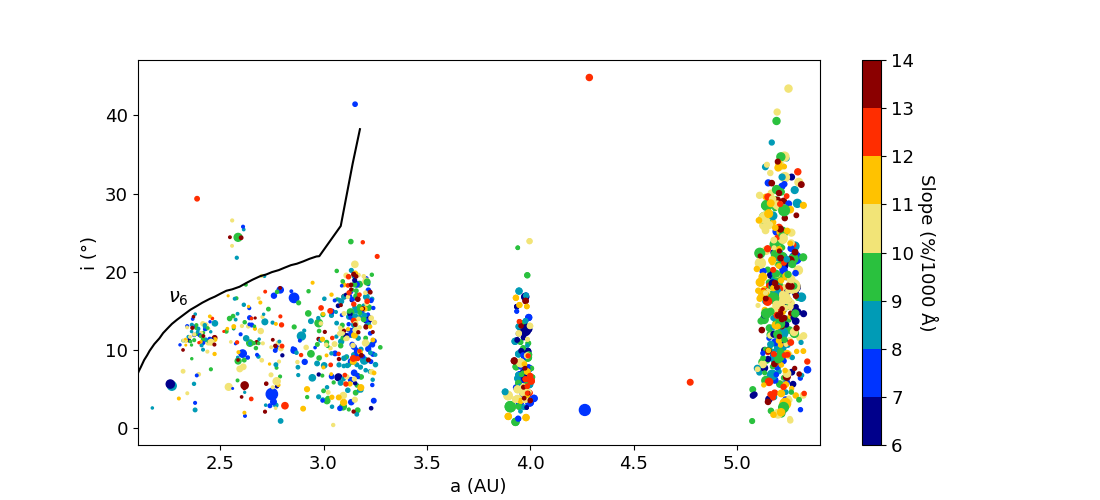}
    \caption{Inclination as a function of the semi-major axis for D- and Z-type objects across the Solar System. The colorbar represents the spectral slope, and the size of the points is proportional to the asteroid's diameter. The black line represents the $\nu_6$ secular resonance \citep{morbidelli2010}.}

   \label{fig:pos}
\end{figure*}

The Trojans have a broader range for inclination, going up to 40$^\circ$, while those in the main belt have mostly i < 20$^\circ$. Trojans D-types are the only population with such an important inclination range \citep{carvano2010}. It is thought to be caused by their capture mechanism in the L4 and L5 swarms of dynamically excited primordial TNOs, yet models have not yet been able to explain this high inclination \citep{bottke2023}. The lack of higher inclination D/Z-types in the main belt is more likely related to resonances, notably the $\nu_6$ one, as noticed also for P-types (Fig. \ref{fig:pos_p} and \ref{fig:pos}).

In Table \ref{tab:SolSys} we report the average values of the slope, albedo, and diameter for the primordial asteroids in the different regions of the Solar System. Asteroids are larger at increasing heliocentric distances, but this is clearly an observational bias. Z-types are significantly larger among Jupiter Trojans than in the Cybele and Hilda, while for D-types, the size range is of the same magnitude in these populations. Conversely, for P-types, the average diameter is larger in the Cybele and Hilda populations than in Jupiter Trojans. \citet{Gartrelle2021b} analyzes $\sim$ 100 D-types, classifies them using the BDM taxonomy, also finding that main belt D-types are about 30\% smaller in size than Trojans. With Gaia's data, we find even a higher difference in size, Non-Trojans D-types, in BDM ($\bar{D}=18.99\pm0.62$ km) are  about 50-60\% smaller on average than Trojans D-types ($\bar{D}=41.84\pm1.82$ km). 

The albedo for both types decreases as the semi-major axis increases, with the exception of Trojans. We observe that D-types in the inner main belt are 25.6\% brighter than those in the outer belt, 32.1\% than in the Cybele, 48.9\%  than the Hilda, and 16.4\% than the Trojans. On the other hand, Z-types in the inner main belt are  24.3\% brighter than those in the outer belt, 38.1\% than the Cybele, 41.3\% than the Hilda, and  35.6\% than the Trojans. That is definitely not an observational bias. An object can get brighter when the surface particle grain sizes are smaller in size. Nonetheless, grain size also impacts the spectral slope with a reddening \citep{cloutis2018}, which is not observed here as the spectral slope does not significantly vary with heliocentric distance. \\
Space weathering may also affect brightness, yet it is poorly constrained for D-type like compositions. Laboratory experiments made on Tagish Lake, and irradiated carbonaceous material show a bluing of the slope and an increase of the albedo \citep{vernazza2013,lantz2017}. If we assume that asteroids located closer to the Sun are more irradiated than the others, this process would explain the anti-correlation between the albedo and the semi-major axis (Table \ref{tab:SolSys}), but we do not observe a spectral bluing. \\
Surface porosity and roughness may also affect the spectral behavior and albedo. \citet{wargnier2025_lab} has studied the effect of different porosities on Phobos' simulants, which are spectrally close to D-types. The authors have shown a brightening as the samples became more porous, yet no significant changes in the spectra in the visible and near-infrared. \\
Moreover, inner main belt asteroids are on average smaller and brighter, which could be attributed to fresher material exposed through collisions as suggested previously.

\begin{figure}
    \centering
    \includegraphics[width=0.9\linewidth]{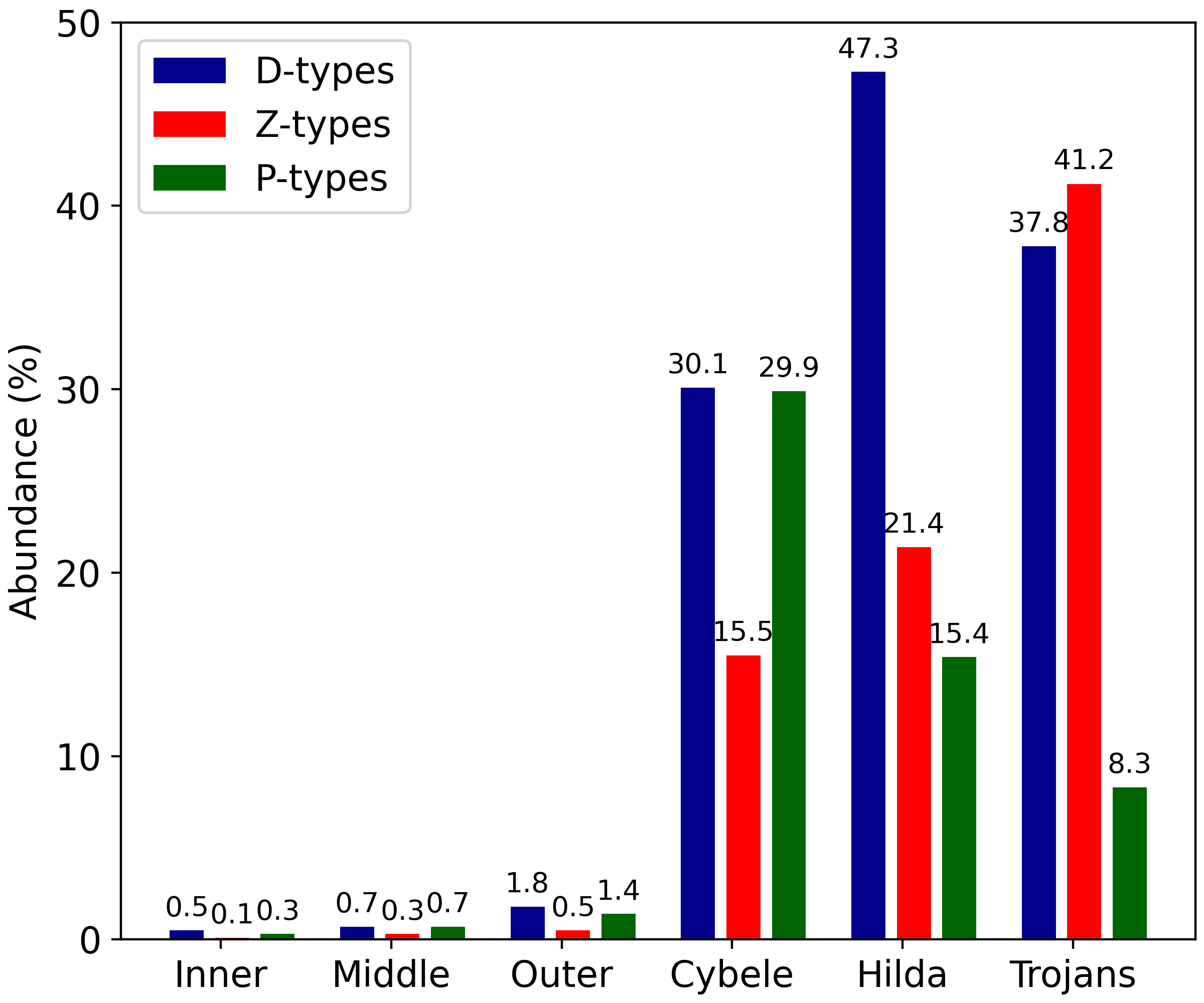}
    \caption{Abundances of D- (in blue), Z- (in red), and P-types (in green), in Mahlke, depending on its position in the Solar System. It was computed from the Gaia's catalog with SNR > 20 of the same category.}
    \label{fig:abun}
\end{figure}

Figure \ref{fig:abun} summarizes the abundances of D-, Z-, and P-types across the Solar System. It was computed considering a sample of asteroids in the Gaia catalog having SNR > 20. We see an increase in abundances with the heliocentric distance: D-types represent 0.5\%, 0.7\%, and 1.8\% of the inner, middle, and outer belt, respectively, while their abundance jumps to 30.1\%, 47.3\%, and 37.8\% for the Cybele, Hilda, and Trojans. On the other hand, Z-types represent 0.1\%, 0.3\%, 0.5\% of the inner, middle, and outer belt, and their proportion drastically increases in the Cybele (15.5\%), Hilda (21.4\%), and Trojans (41.2\%). As expected, the farther away from the Sun, the more abundant the Z-types become. Surprisingly, that is not true for D-types, they are more abundant in the Hilda than the Trojans. P-types are more peculiar, there are 0.3\%, 0.7\%, and 1.4\% of P-types in the inner, middle, and outer main belt. They also represent 29.9\%, 15.4\%, and 8.3\% of the Cybele, Hilda, and Trojans. Nonetheless, the three types have an abundance jump at the Cybele region, which can be explained by the difficulty for asteroids to cross the 2:1 MMR. 

\begin{table*}
    \caption{Table referencing the mean slope, albedo, diameter, and abundance of D- and Z-types in Mahlke's taxonomy, according to their position in the Solar System. }
    \begin{tabular}{c c c c c c c}
    \hline
      & Inner & Middle & Outer & Cybele & Hilda & Trojans\\
    \hline
    \multicolumn{7}{c}{D-types}\\
    \hline
    $\bar{S} $(\%/1000~\AA) & $9.34\pm0.22$& $9.27\pm0.18$ & $8.78\pm0.11$& $7.99\pm0.11$&$8.05\pm0.15$&$8.74\pm0.14$\\
    $\bar{p_v} $(\%) & $9.08\pm0.23$ & $8.40\pm0.25$&$7.23\pm0.17$&$6.87\pm0.27$&$6.10\pm0.19$&$7.80\pm0.17$\\
    $\bar{D} $(km) &$9.29\pm1.25$&$13.
    77\pm1.84$ &$16.89\pm0.90$& $26.05\pm3.16$&$31.22\pm2.29$&$33.27\pm1.34$\\
    \hline 
    \multicolumn{7}{c}{Z-types}\\
    \hline
    $\bar{S} $(\%/1000~\AA) &$12.09\pm0.44$&$12.11\pm0.36$&$11.41\pm0.19$&$11.84\pm0.34$&$12.01\pm0.25$&$11.37\pm0.14$\\
    $\bar{p_v} $(\%) &$9.25\pm0.59$&$8.56\pm0.33$&$7.44\pm0.23$&$6.70\pm0.38$&$6.54\pm0.32$&$6.82\pm0.12$\\
    $\bar{D} $(km) & $9.68\pm2.10$ &$13.69\pm2.20$&$16.99\pm0.91$&$27.87\pm30.4$&$25.43\pm1.87$&$43.27\pm1.88$\\
    \hline
    \multicolumn{7}{c}{P-types}\\
    \hline
    $\bar{S} $(\%/1000~\AA) &$2.99\pm0.12$ &$2.93\pm0.08$ & $2.79\pm0.05$& $3.09\pm0.17$& $3.51\pm0.26$& $4.54\pm0.26$\\
    $\bar{p_v} $(\%) &$5.50\pm0.13$ &$5.53\pm0.09$ & $5.44\pm0.08$& $4.79\pm0.18$ & $4.72\pm0.26$& $7.24\pm0.40$\\
    $\bar{D} $(km) &  $11.44\pm2.21$ & $21.16\pm3.21$&$29.76\pm2.80$ &$60.54\pm8.19$ & $55.24\pm9.22$ & $40.72\pm5.30$\\
    \hline
    \end{tabular}
    \label{tab:SolSys}
\end{table*}

Using only the D- and Z-types in the Trojans from \citet{Fornasier2025}, we re-computed the correlation coefficients for Trojans. They are shown in Table \ref{tab:corr_tro}. Trojans, as observed for primordial MBAs, show a weak anti-correlation between the albedo and the diameter, but stronger. It would mean that the bigger an asteroid is, the darker it is. It seems to be, for Trojans, an observational bias, as it is easier to detect dark asteroids when they are bigger. The Trojan D-types show an anti-correlation between the albedo and the inclination, meaning the brighter they are, the less inclined they are. However, this anti-correlation is weak and with a low significance p-value. \\
Z-types have an anti-correlation between $D$ and $p_v$, which is also observed in MBA. In addition, Trojan Z-types present an anti-correlation between the size and the slope. Hence, the larger the asteroids are, the less red they are. It is also noticed in the main belt, and in the  Cybele and Hilda populations. 

\begin{table}
    \centering
    \caption{Pearson's correlation coefficient and p-value between different parameters for D- and Z-types in the Trojans (with no distinctions between the L4 and L5 swarms).}
    \begin{tabular}{c c}
    \hline
       Variables  & Correlation coefficient ($P_r$)  \\
       \hline 
       \multicolumn{2}{c}{D-types}\\
       \hline
        $p_v$ vs $i$ & -0.22 (0.002)\\
        $p_v$ vs $D$ & -0.47 ($1.2 \times 10^{-9}$)\\
        \hline
        \multicolumn{2}{c}{Z-types}\\
        \hline
        $\bar{S}$ vs $D$ & -0.40 ($3.5 \times 10^{-9}$)\\
        $p_v$ vs $D$ & -0.41 ($1.0\times10^{-9}$)\\
        \hline
    \end{tabular}
    \label{tab:corr_tro}
\end{table}

We do not see spectral slope's variations at different heliocentric distances for a given spectral type. However we are limited by the visible range of Gaia's spectra. For example, \citet{Gartrelle2021b} reported increasing VNIR spectral slope (in the 1.5 -- 1.8 $\mu$m and 2.0 -- 2.45 $\mu$m region) as D-types get closer to the Sun. They attributed this trend to different processes such as space weathering, collisions and YORP spin-up for the smaller bodies. This study is not the only one reporting differences in the infrared, despite the same D-type classification. \citet{oriel_2024_red} separated red asteroids into two groups using infrared spectra with moderate and steep slopes, similarly to the red and less-red asteroids in the Trojans defined by \citet{emery2011a}. It could suggest another source of main-belt D-types within Jupiter's orbit \citep{oriel_2024_red}. In another study in the mid infrared (5 -- 40 $\mu$m), \citet{oriel_2024_prim} studied the difference between spectra of D- and P-types in the main belt and in the Trojans. The authors reported that the majority showed similar features, no matter the position in the Solar System, yet a small part exhibited differences in the 10 $\mu$m feature. Therefore, similar spectra in the VNIR do not necessarily reflect a similar composition and origin. It suggests that from those spectral differences in the VNIR, D-types could have different origins \citep{oriel_2024_red,Gartrelle2021a}.Concerning the Gaia spectral data here analyzed, we are forcedly limited in the compositional interpretation because we only have spectrophotometric data in the visible range.  

\section{Discussion}
\subsection{Implantation of TNOs in the main belt}

As seen previously, primordial D-type and Z-type abundances increase beyond the main belt, and they dominate the Trojan swarms.   
It is thought that Trojans are escaped TNOs \citep{morbidelli2005, emery2011a}, captured by Jupiter during the planetary migration phase. \citet{Fornasier2025} noticed that the Trojan population lacks extremely red objects observed in the transneptunian region. They explained this lack of extremely red objects by the removal of the organic, red-rich crust through sublimation when TNOs were injected at closer heliocentric distances and captured by Jupiter. Moreover, \citet{levison2009} have shown that some TNOs might have been implanted in the main belt, and may be at the origin of the observed D-types. It was suggested that D-types share the same origin as the Trojans, i.e. they come from the Kuiper belt region, and progressively moved up to the inner main belt through collisions and Yarkovsky effects \citep{DeMeoDtype}. This explanation would explain the lack of large D-types as we get closer to the inner belt.  \\
\citet{DeMeoDtype} considered a first implantation of D-types in the MMB, then a further implantation in the IMB crossing the 3:1 MMR with the energy given by a catastrophic collision. If this hypothesis is correct, we would expect D/Z-type families in the inner and middle belt, which is not observed. Hence, from the absence of D-type families in the main belt, we think it is unlikely that they crossed the MMR through collisions. Nonetheless, if a catastrophic collision took place before the main belt and their implantation, the lack of families is expected. On the other hand, we noticed several P-type members of primitive families, suggesting that they likely formed in-situ, contrary to D-types. \\
The second hypothesis given by \citet{DeMeoDtype} is that D-type asteroids were implanted from the outer to the inner main belt by Yarkovsky effect, which affects  bodies smaller than 30 km. This explanation is in agreement with the  anti-correlation we have found between $a$ and $D$. \citet{volkrou2016} evokes the possibility of the Yarkovsky effect drifting smaller bodies toward the MMR. Nevertheless, the authors noted that this can be true only at small heliocentric distances, hence not likely after the outer main belt. 

The study of the Tarda and Tagish Lake meteorites, the former of which exhibits a spectrum closer to P-types and the latter to D-types, shows a possible common origin from the same parent body, as they share similarities in their petrologies \citep{schrader2024}. The authors therefore suggest that P- and D-types originate from the same object and that their spectral differences are caused by slight differences in composition and/or grain size. Nevertheless, P-types are generally larger than D-types (see Table \ref{tab:SolSys}). This trend is consistent throughout the main belt, Cybele and Hilda, which could suggest that P-types are composed of more robust material \citep{dahlgren1997,lagerkvist2005}. In addition, P-type members were found within primitive families dominated by mostly C- and X-types in BDM taxonomy, suggesting that they may represent a space weathering stage of surface evolution of dark and carbon-rich asteroids. Therefore, it seems unlikely that P- and D- types come from the same parent bodies that originated from the outer solar system.

\citet{volkrou2016} showed that dynamical simulations including a fifth giant planet (MN12) made it possible to implant P- and D-types in the inner and middle main belt regions from the trans-Neptunian region. The initial conditions needed are the presence of the fifth giant planet and the presence of the primordial reservoir at 23 -- 30 AU. Their simulations give an overabundance of D/P-types with diameters above 10 km, and they have predicted the presence of $\sim$140-280 D/P-types in the middle belt and 8000-24000 in the outer one. When looking at the D-, Z-, and P-types as a whole, with $D > 10$ km, we find 135 objects in the middle belt, which is consistent with the findings of \citet{volkrou2016}, and 300 in the outer one, confirming that they are more numerous in the outer main belt than in the inner, but with abundances that are much smaller than predicted by \citet{volkrou2016}. In these simulations, collisions and Yarkovsky are not needed to implant these objects across the main belt, and are not taken into account in their simulations. Nonetheless, these processes imply a loss of objects and may potentially explain the difference between the observed number of D- and P-types in the outer belt and the simulated one. 

In the Cybele and Hilda, P- and D-types dominate, but larger bodies belong to P-types, especially compared to D-types. This result could suggest a formation in-situ for P-types. \\
In the main belt, P-types are sometimes found in C/X- families. The difference in slope between a slightly red C-type and a moderately sloped P-type is tenuous, and there may be spectral slope evolution related to SW processes. Space weathering on C-types is likely dependent on the original albedo, if low ( $<\sim5-6$\%) they become brighter and bluer. On the contrary, if their albedo is high ($> \sim7-9$\%) they redden and become darker \citep{lantz2017}. 

\subsection{Comparison to comets}

Some of the objects in our dataset displayed high inclinations and eccentricities, closer to cometary orbits. Hence, we have computed the Tisserand parameter, $T_J$, as follows: $$T_J=\frac{a_J}{a}+2\cos{i}\sqrt{(1-e)^2\frac{a_J}{a}}$$ where $a_j$ is Jupiter's semi-major axis, $a$ the semi-major axis of the object, $e$ its eccentricity and $i$ its inclination. Objects with $T_J$ $\leq$ 3 are dynamically closer to Jupiter family comets (JFC), while objects with $T_J$ > 3 are closer to asteroids \citep{tisserand}. We have identified 42 D- and 18 Z-types with $T_J$  < 3.  The fact that activity has not been observed for these bodies places them in the Asteroids on Cometary Orbits (ACO), which could be extinct comets. However, it is important to note that the distinction between comets and asteroids is not often clear, and cometary activities have short lifetimes and can not be caught during a given observation \citep{Fornasier2021}. The continuum between the two types of small bodies is debated \citep{Jewitt22}. ACOs typically show primitive spectra, and their spectra cannot differentiate between an asteroidal or cometary origin \citep{licandro2008}. One of the most known D-type ACOs is (944) Hidalgo \citep{hargrove2008}, which has $a=5.7$ AU, $i=50.9^\circ$, and $e=0.667$.  

Within the main belt (2.0 -- 3.3 AU), there is only one D-type that has $T_J$ below 3 : (2938) Hopi, and no Z-types. Hopi has a  3.15 AU semi-major axis, an eccentricity of 0.13, and an inclination of 41.43° which for main belt objects is uncommon. There is also one P-type: (71572) 2000 DW42 in the outer main belt. \citet{Hsieh2016} have shown the possibility for JFC to get implanted in the main belt and enter stable orbits, even during present days and not necessarily during giant planets' migrations. In their simulations, the Tisserand parameter raised above 3, with a maximum of 3.05. It could suggest that a small portion of D-types found in the outer main belt, with high inclinations and eccentricities, are extinct comets, even within the asteroids with $3.0 < T_j < 3.05$. 

\begin{figure}
    \centering
    \includegraphics[width=\linewidth]{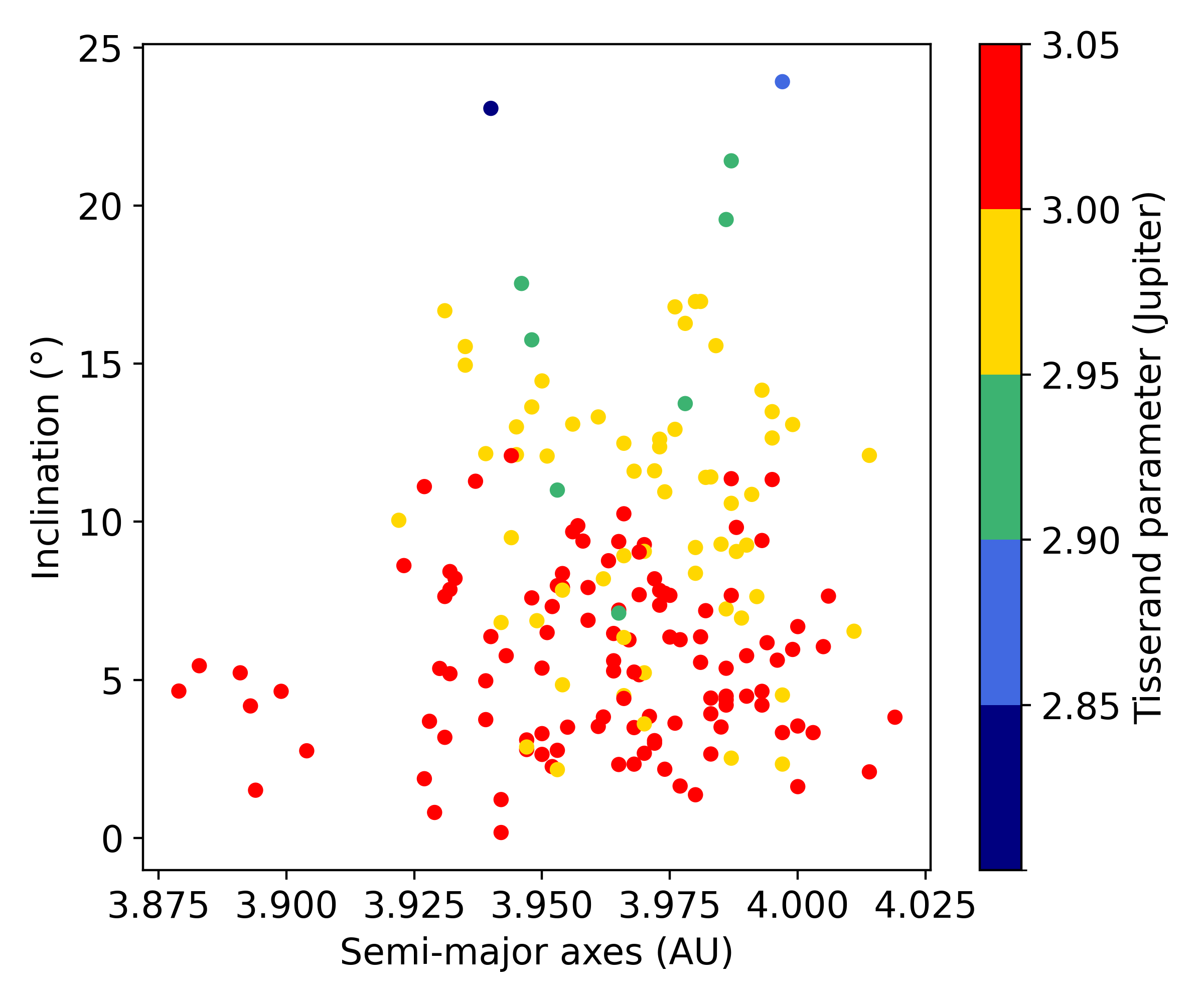}
    \caption{Inclination versus semi-major axis of the Hilda population. The colorbar represents the Tisserand parameter value. }
    \label{fig:tiss_pos}
\end{figure}

Eight objects in the Cybele group are dynamically closer to the JFC (4\% of the Cybele population). As for the Hilda, 28\% of the objects have $T_J$ < 3 (68 asteroids out of 180). The objects in the Cybele and Hilda with the lowest $T_J$ are P-, D-, and Z-types, that is objects with spectral slopes commonly observed for cometary nuclei \citep{Filacchione_2024}. The Tisserand parameters and their positions are shown in Figure~\ref{fig:tiss_pos}. 
\\Close to the Hilda population, there is the quasi-Hilda comets (QHC) region, which are unstable. They are thought to be JFC that moved out of Jupiter's orbit \citep{kresak79}. Furthermore, a significant proportion of the Hilda are in the JFC zone. Those QHC contain some escaped D-type Hilda \citep{disisto2005}, and they originated from Centaur or  transneptunian, implanted there during giants planets' perturbations. Some QHC exhibit cometary activity \citep{correaotto24, gilhutton2016}. Considering their proximity, it is not surprising to see as many ACOs in the Hilda as it is a known stable region that might have captured some QHC. 

In total, there are 80 ACOs in the dataset, taking into consideration both Cybele and Hilda. They are taxonomically diverse, and are mostly primitive : D (52.5\%), Z (22.5\%), P (8.8\%), M (5.0\%), P/D (3.8\%), X (2.5\%), Ch (2.5\%), S (1.3\%), and C (1.3\%). The presence of one S-type (asteroid 692) and four M-types (asteroids 1371, 11249, 17212, and 88237) is surprising. 

\citet{licandro2008} in a spectroscopic survey of 24 ACOs reported a strong anti-correlation between the Tisserand parameter and the slope in the visible. Hence, we computed this correlation for the ACOs present in our dataset. We do find a weak anti-correlation with a coefficient of -0.22 and a low p-value (0.046). Moreover, this anti-correlation is recovered with a coefficient of -0.24 ($P_r=0.026$) for P-type family members. \citet{licandro2008} also report a weak correlation between the slope and $a$, as well as between the slope and the eccentricity, which are also observed in our dataset (the coefficients are 0.37 and 0.23, respectively) with low p-values (0.00068 and 0.39, respectively). Moreover, there is a moderate anti-correlation between the semi-major axis and the inclination (-0.52, $P_r=7.8\times10^{-7}$), meaning the farther away from the Sun they are, the less inclined they tend to be. 

\begin{figure}
    \centering
    \includegraphics[width=0.5\textwidth]{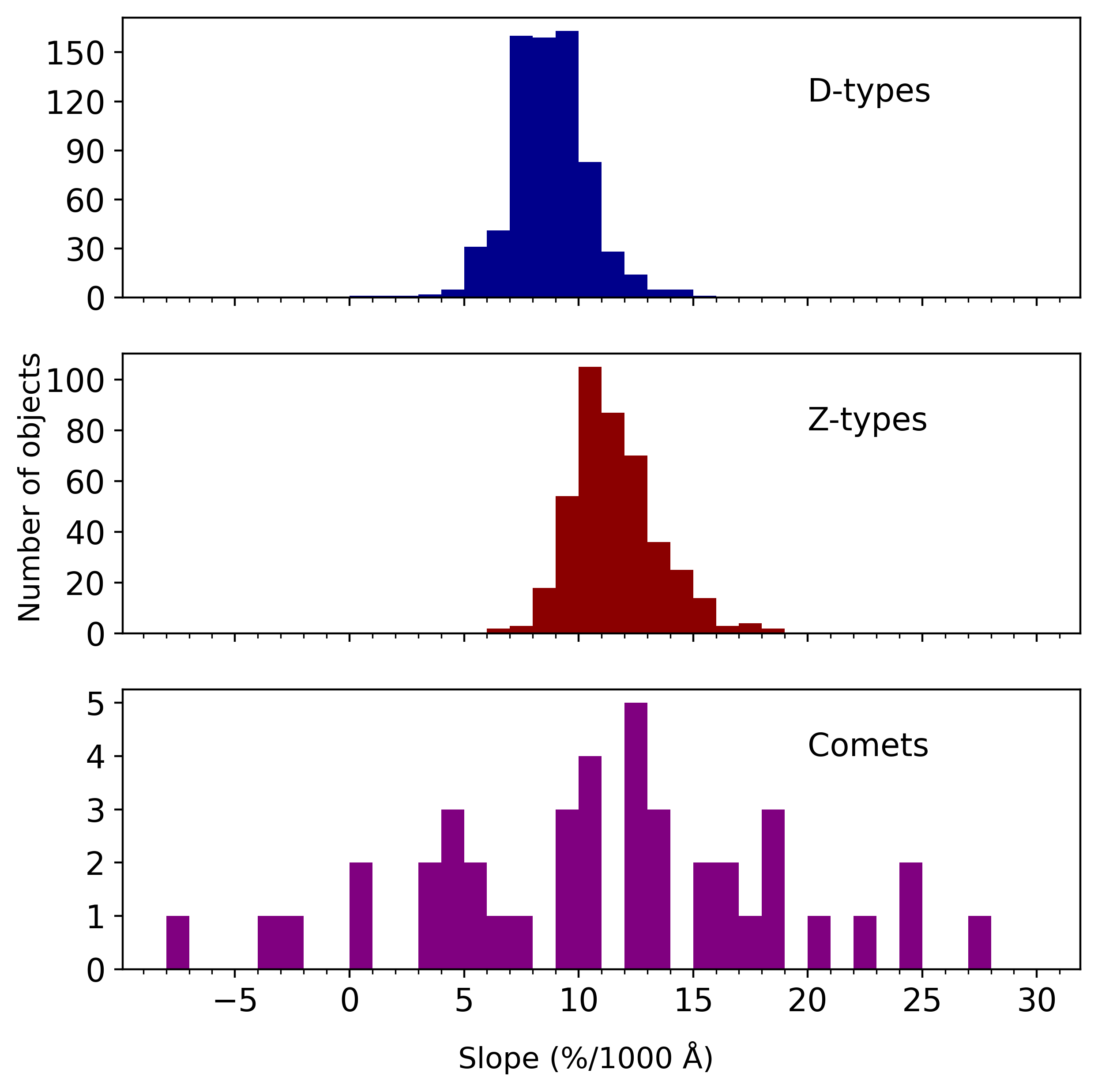}
    \caption{Slope distribution of the whole D-type population (top panel), and Z-types (middle panel) here studied, and comets (bottom one). The slope has a bin of 1 \%/1000~\AA~ bin.}
    \label{fig:slope_dist_comets}
\end{figure}

We have compared (Figure \ref{fig:slope_dist_comets}) the slope distribution of all the D- and Z-types in our dataset, including Trojans,  with that of comets from the Minor Bodies in the Outer Solar System (MBOSS) catalog \citep{hainaut2012}. There are only 48 comets in this catalog, much less compared to the primordial asteroids here investigated. As can be seen,  comets have spectral slopes covering a broader range, from -8 up to 28 \%/1000~\AA. However, it should be noted that the color of comets often includes the contribution of the coma, which has different effects depending on the composition, particle size, and geometric conditions of the observations. Furthermore, Rosetta's observations of comet 67P/CG revealed that the spectral slope/colors change depending on the season, heliocentric distance, and rotational phase, related to diurnal and seasonal cycles of water and to the activity. A less steep slope is observed close to perihelion, when the dusty, organic-rich mantle is lifted by intense activity, while red spectra are observed in comet 67P when activity decreases at larger heliocentric distances \citep{Fornasier_2015, Filacchione_2024,desanctis2015}.

\subsection{Comparison with the Martian moons}
\begin{figure}
    \centering
    \includegraphics[width=0.4\textwidth]{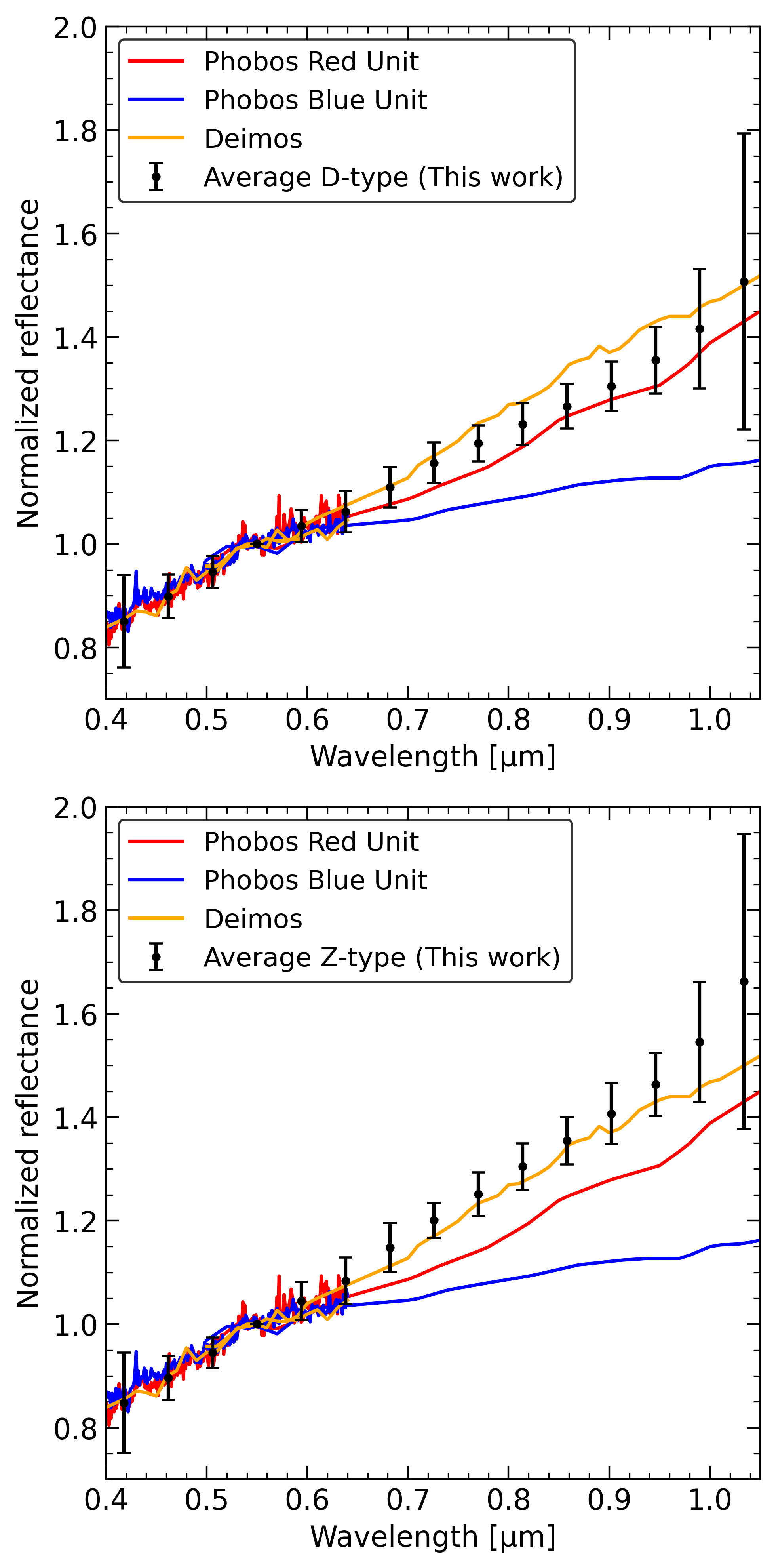}
    \caption{Comparison of the spectra of Phobos and Deimos with the average spectra of D- and Z-type asteroids derived from Gaia data. The Phobos and Deimos spectra were retrieved from \cite{Fraeman_2012} for the visible-near-infrared wavelength range and from \cite{Mason_2023} for the ultraviolet-visible range. All the spectra were normalized at 550 nm.}
    \label{fig:comparison_martian_moons}
\end{figure}
The two Martian moons, Phobos and Deimos, are characterized by their peculiar shape, orbital, and spectrophotometric properties. Both are small bodies with irregular shapes and exhibit a diameter of 26 $\times$ 24 $\times$ 18 km for Phobos, and 15 $\times$ 12 $\times$ 10 km for Deimos \citep{Thomas_1989}. They present particularly circular ($<0.02$, \citealt{Jacobson_2010}), low-inclination orbits ($<2^\circ$, \citealt{Jacobson_2010}). Phobos is the main target of the JAXA Martian Moons Exploration (MMX) mission, which will collect and bring back to Earth samples of Phobos. The spectra of Phobos and Deimos have been found to be particularly close to primitive asteroids such as D-types (e.g., \citep{Rivkin_2002, Fraeman_2012, Takir_2022}), and more recently Z-types \citep{wargnier_2025_spec}. The photometric properties of the two moons have also been shown to be similar to primitive objects, including the 67P/Churyumov–Gerasimenko cometary nucleus \citep{Fornasier_2024, wargnier_2025_deimos}. These various properties have led to two main hypotheses to explain the origins of the Martian moons. The first one is the giant impact theory in which an object impacts Mars in its early history creating a debris disk, that re-accretes into Phobos and Deimos \citep{Burns_1992, Craddock_2011, Rosenblatt_2011}. This theory explains well the orbital properties of the Martian moons but fails to explain the spectroscopic and photometric properties.  In fact, the spectrophotometric properties of the moons are very different from those observed for the different Martian terrains or Martian meteorites. The second hypothesis claims that Phobos and Deimos are captured primordial asteroids, which explains their spectroscopic properties similar to D/Z-types \citep{Hunten_1979, Higuchi_2017}. However, the circular and low-inclined orbits observed today cannot be so far reproduced by dynamical models in the case of captured asteroids. Recent alternative theories offer a possible explanation to reconcile the orbital and spectrophotometric properties, by invoking the partial disruptive capture of a single asteroid or of a bilobated comet \citep{Kegerreis_2025, Fornasier_2024}. \par
To help interpreting Martian moons origin, we also compared the spectra of Phobos and Deimos with the average Gaia spectra of the D- and Z-type asteroids here studied (Fig. \ref{fig:comparison_martian_moons}). While the average D-type exhibits a 0.55 -- 0.814 µm spectral slope of 8.78 $\pm$ 0.29 \%/1000~\AA, the Z-types are obviously redder with a spectral slope of 11.54 $\pm$ 0.39 \%/1000~\AA. The Phobos Blue Unit (BU) is significantly less red than D- or Z-types from Gaia, we computed a spectral slope of 3.63 $\pm$ 0.22, from the combined spectra from \citet{Fraeman_2012} and \citet{Mason_2023}. Similarly, the Phobos Red Unit (RU) is slightly less red (7.32 $\pm$ 0.44 \%/1000~\AA) than Deimos (10.43 $\pm$ 0.63 \%/1000~\AA) in the 0.55 -- 0.814 µm wavelength range. We observed that Deimos is closer to Z-types than D-types, while the Phobos RU is closer to D-types. This result slightly differs from \cite{wargnier_2025_spec}, which shows that both Phobos and Deimos are overall more consistent with Z-type asteroids. One reason could be that, we used here only the visible wavelength range, which necessarily limits the comparison. Anyway, this work shows that primordial asteroids similar to the Martian moon's spectra exist not only in the outer solar system, but in the main belt, even if in low amount. The detection of several very red objects in the entire main belt supports the possibility of a capture scenario not only from the Jupiter Trojans or the Hilda asteroids, but also from the inner main belt, as stated in \cite{wargnier_2025_spec}.

\section{Conclusions}

We have classified and analyzed 318 D-types, 124 Z-types, and 265 P-types in the main belt, and classified Cybele and Hilda asteroids present in Gaia DR3. The main results of this analysis are:

\begin{itemize}

    \item Primordial D- and Z-types are present in the main belt, even if in small amounts: 0.5/0.1\% in the IMB, 0.7/0.3\% in the MMB, and 1.8/0.5\% in the OMB for D-/Z-types, respectively. Their abundances suddenly increase after the 2:1 MMR to 30.1/15.5\% in the Cybele, 47.3/21.4\% in the Hilda and 37.8/41.2\% in the Trojans, for D-/Z-types, respectively. Some D-types in the main belt have TNO-like spectral behavior, such as (269) Justitia which shows a very red spectrum. A few present a spectral flattening beyond 0.8 $\mu$m, similarly to what is observed on some cold classical TNOs. All this suggests that D- and Z-types likely originated from the outer Solar System.
    
    \item We have classified 193 Cybele from Gaia DR3, and 180 Hilda. Cybele is dominated by P- and D-types, while Hilda by D- and Z-types. Their spectral slope distributions are bimodal. D-types are, on average, smaller than P-types. This can indicate the presence of a more robust material in P-types. The higher abundance of P-types, compared to D- and Z-types in the main belt and Cybele region, and its drastic decrease after 4 AU indicates that P-types likely formed "in situ".

    \item The diameter of D- and Z-types in the main belt is weakly anti-correlated to the semi-major axis in agreement with previous studies \citep{DeMeoDtype,Gartrelle2021b}. If the absence of small D-types at large $a$ is simply an observational bias, the lack of large D-types in the IMB is real. The inner belt D/Z-types are also brighter than those located farther. This may indicate that rejuvenating processes were more effective close to the Sun (collisions and/or space weathering effects).

    \item An important portion of P-types in the main belt belong to C- and X-type families (35\% of them), potentially hinting at a common origin between C- and P-types.
    
    \item Several primordial asteroids, especially in the Hilda, have a Tisserand parameter lower than 3. Considering the resemblance between primordial asteroids' spectra and comets' nuclei, it could hint that certain primordial asteroids could have cometary origins.
    
    \item The presence of D- and Z-types in the main belt, coupled with the spectral similarities of Phobos and Deimos to primordial red asteroids, further reinforce the hypothesis that the two Martian moons can be captured asteroids. 

\end{itemize}

\section*{Acknowledgements}
This work has received support from France 2030 through the project Académie Spatiale d'Île-de-France\\ (https://academiespatiale.fr/), managed by the National Research Agency under the reference ANR-23-CMAS-0041, and the Centre National d’Etude Spatial (CNES). 
This work has made use of data from the European Space Agency (ESA) mission
{\it Gaia} (\url{https://www.cosmos.esa.int/gaia}), processed by the {\it Gaia}
Data Processing and Analysis Consortium (DPAC,
\url{https://www.cosmos.esa.int/web/gaia/dpac/consortium}). Funding for the DPAC
has been provided by national institutions, in particular the institutions
participating in the {\it Gaia} Multilateral Agreement. This work is based on data provided by the Minor Planet Physical Properties Catalogue (MP3C) of the Observatoire de la Côte d'Azur. This project was supported by the French Planetology National Program (INSU-PNP). 

\printcredits
\bibliographystyle{cas-model2-names}
\bibliography{ref}

\newpage
\appendix
\onecolumn
\section{Additional figures}
\subsection{Main Belt}
\subsubsection{P-types}
\begin{figure*}[!htbp]
    \centering
    \includegraphics[width=0.9\textwidth]{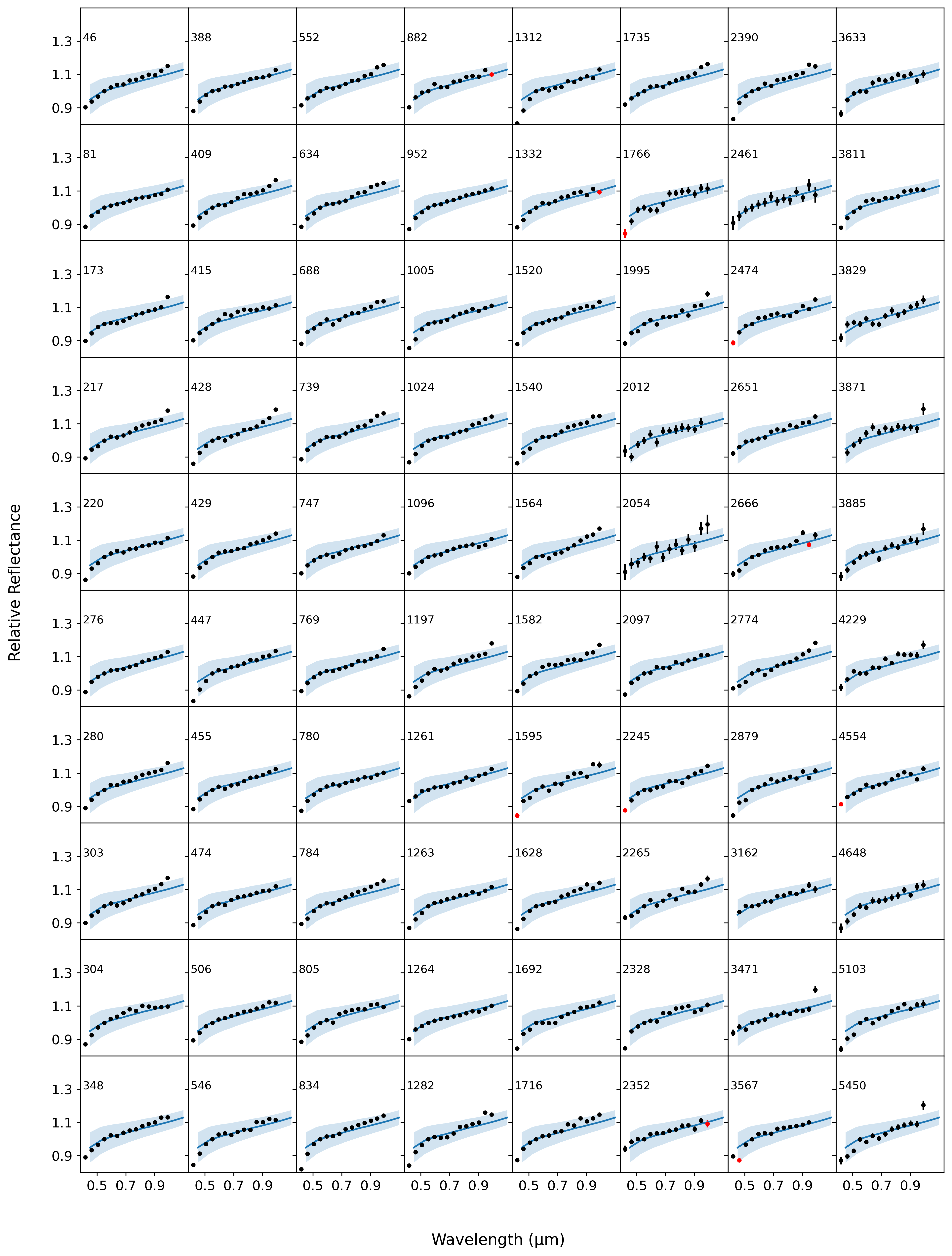}
    \caption{Main belt P-types spectrophotometry from Gaia DR3 catalog (the black dots), with superimposed the best fit class from Mahlke taxonomy (blue line) \citep{Mahlke_2022}. The red dots represent data to be considered with caution (flagged 1 in DR3 catalog).} 
\end{figure*}

\begin{figure*}[!htbp]
    \centering
    \includegraphics[width=0.9\textwidth]{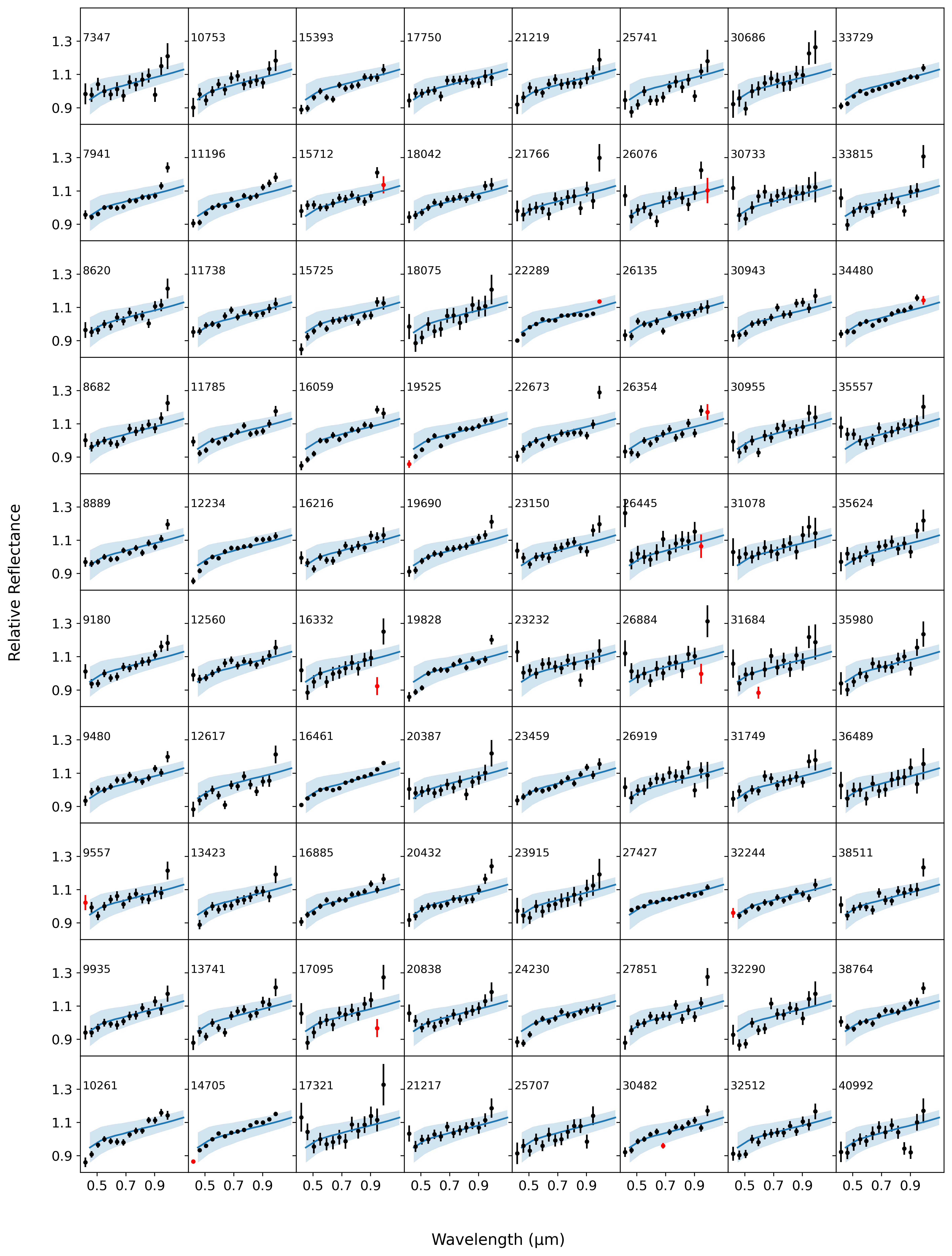}
    \caption{Main belt P-types spectrophotometry from Gaia DR3 catalog (the black dots), with superimposed the best fit class from Mahlke taxonomy (blue line) \citep{Mahlke_2022}. The red dots represent data to be considered with caution (flagged 1 in DR3 catalog).} 
\end{figure*}
\begin{figure*}[!htbp]
    \centering
    \includegraphics[width=0.5\textwidth]{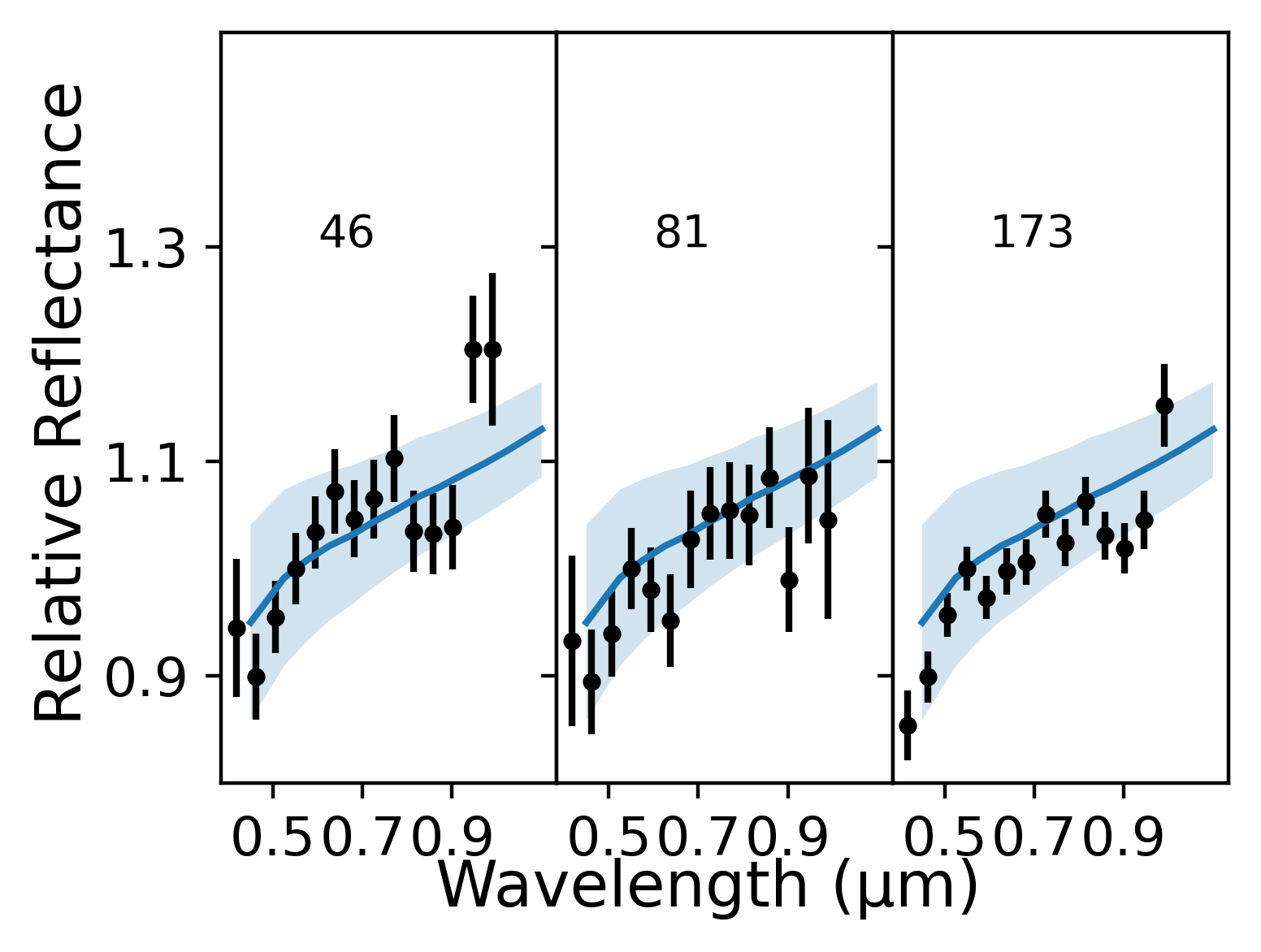}
    \caption{Main belt P-types spectrophotometry from Gaia DR3 catalog (the black dots), with superimposed the best fit class from Mahlke taxonomy (blue line) \citep{Mahlke_2022}. The red dots represent data to be considered with caution (flagged 1 in DR3 catalog).} 
\end{figure*}

\clearpage
\subsubsection{D-types}
\begin{figure*}[!htbp]
    \centering
    \includegraphics[width=0.9\textwidth]{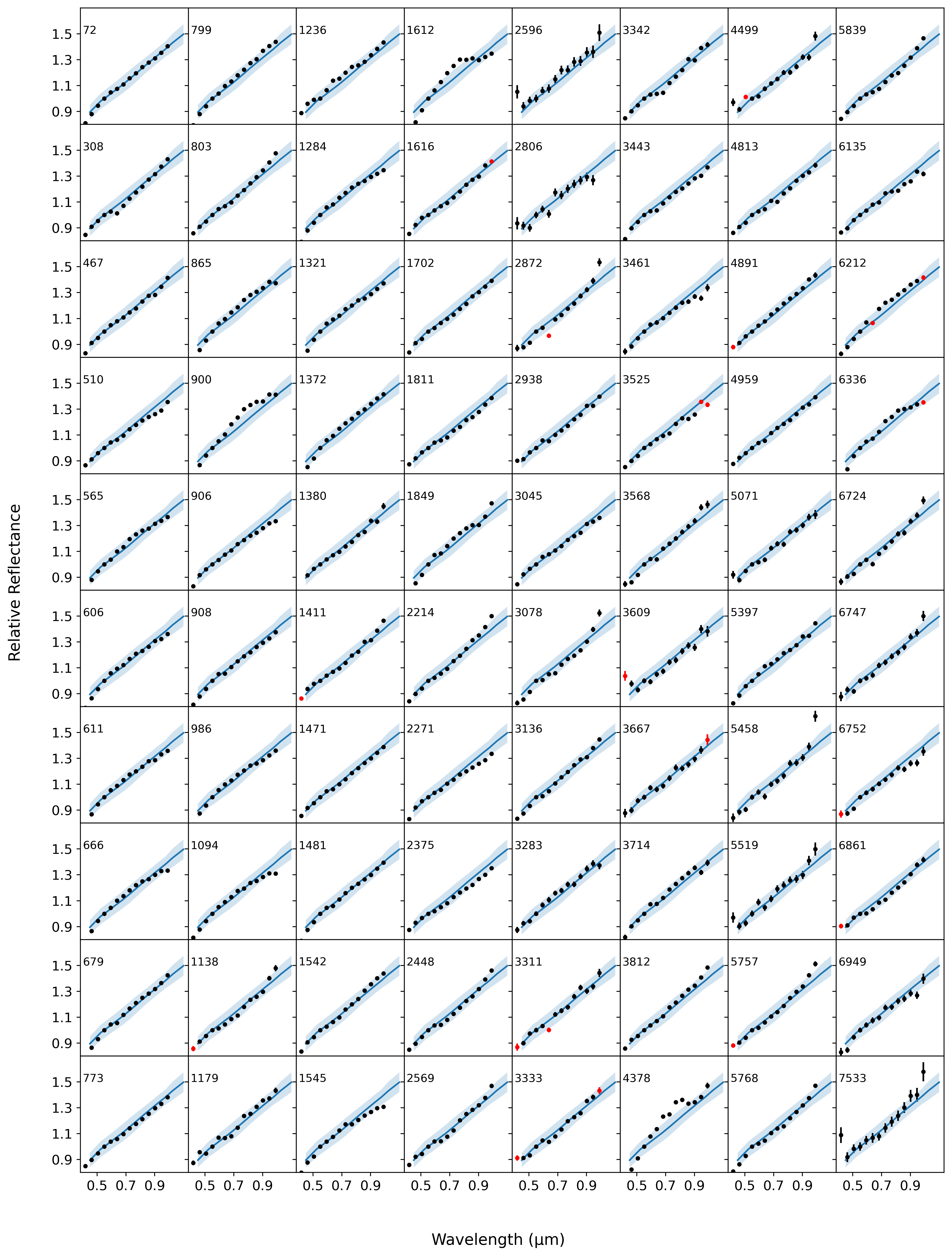}
    \caption{Main belt D-types spectrophotometry from Gaia DR3 catalog (the black dots), with superimposed the best fit class from Mahlke taxonomy (blue line) \citep{Mahlke_2022}. The red dots represent data to be considered with caution (flagged 1 in DR3 catalog).} 
\end{figure*} 

\begin{figure*}[!htbp]
    \centering
    \includegraphics[width=0.9\textwidth]{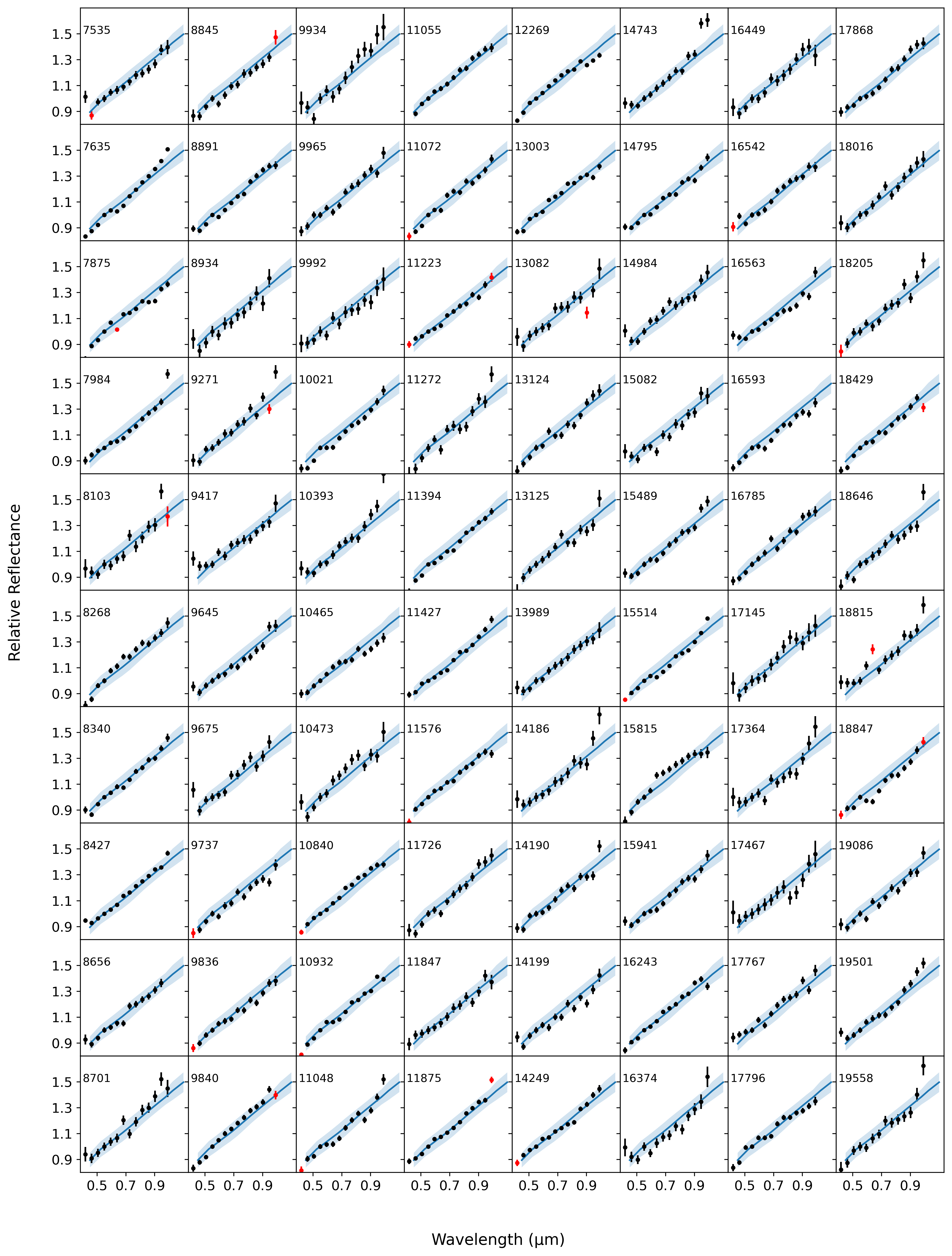}
    \caption{Main belt D-types spectrophotometry from Gaia DR3 catalog (the black dots), with superimposed the best fit class from Mahlke taxonomy (blue line) \citep{Mahlke_2022}. The red dots represent data to be considered with caution (flagged 1 in DR3 catalog).} 
\end{figure*}

\begin{figure*}[!htbp]
    \centering
    \includegraphics[width=0.9\textwidth]{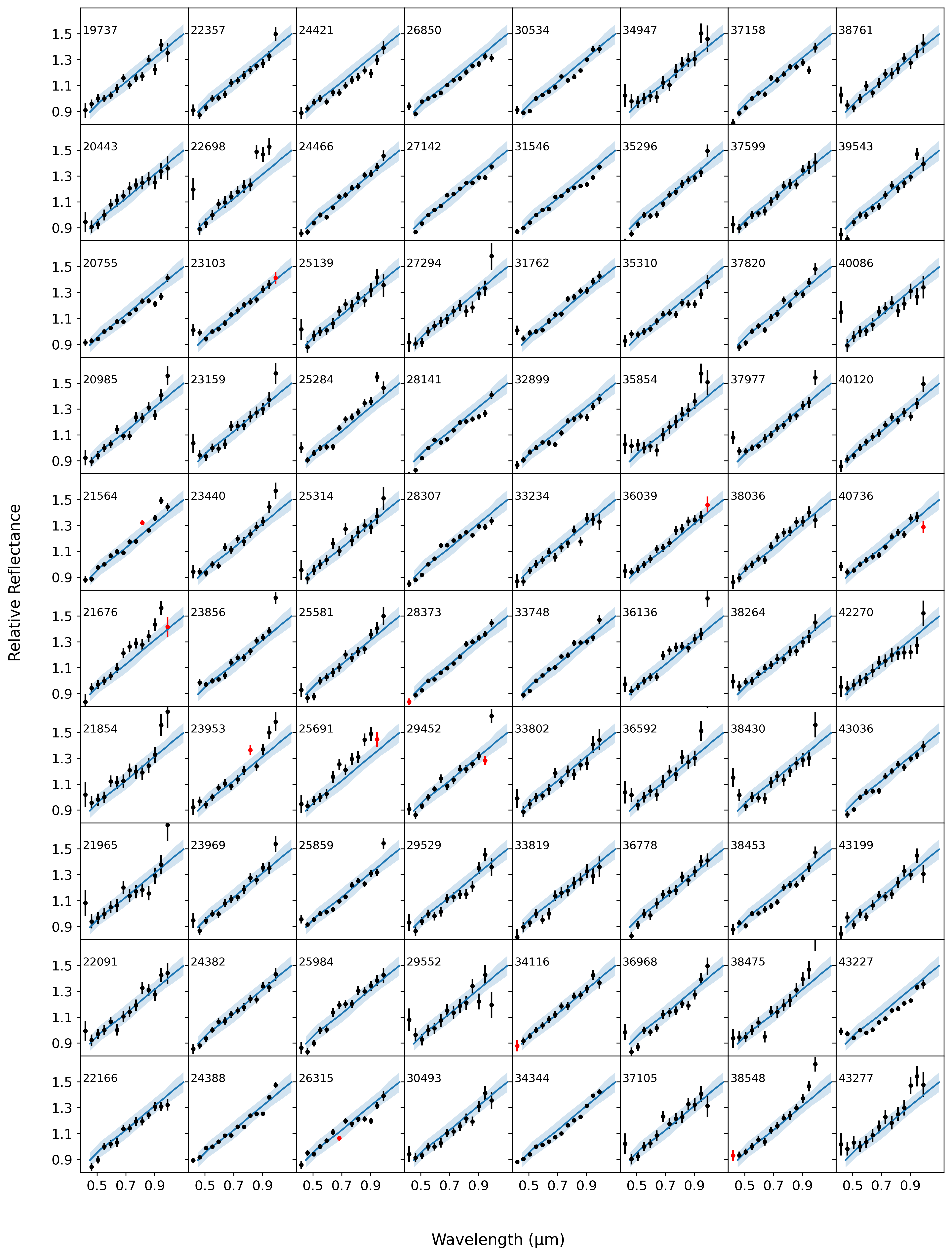}
    \caption{Main belt D-types spectrophotometry from Gaia DR3 catalog (the black dots), with superimposed the best fit class from Mahlke taxonomy (blue line) \citep{Mahlke_2022}. The red dots represent data to be considered with caution (flagged 1 in DR3 catalog).} 
\end{figure*}

\begin{figure*}[!htbp]
    \centering
    \includegraphics[width=0.9\textwidth]{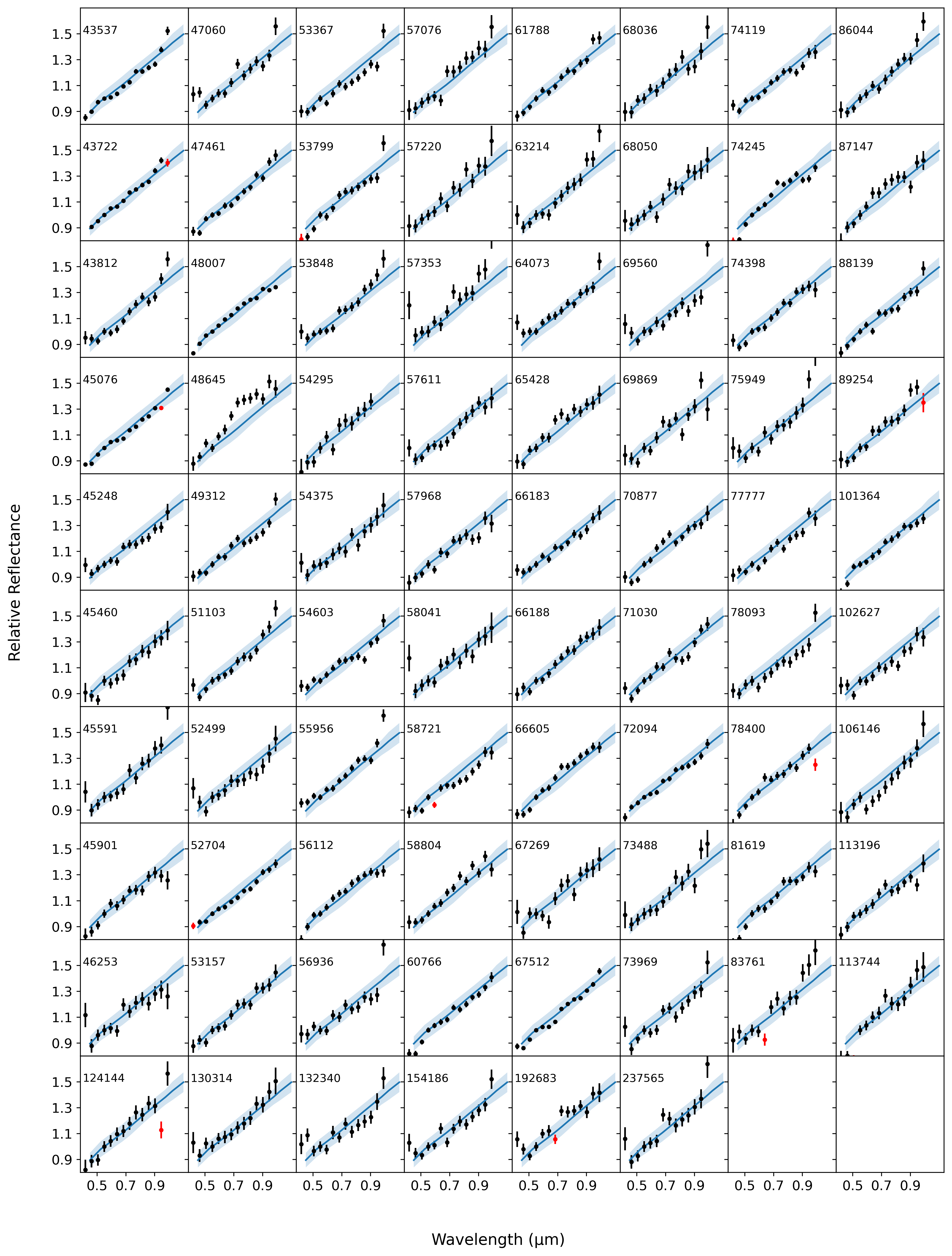}
    \caption{Main belt D-types spectrophotometry from Gaia DR3 catalog (the black dots), with superimposed the best fit class from Mahlke taxonomy (blue line) \citep{Mahlke_2022}. The red dots represent data to be considered with caution (flagged 1 in DR3 catalog).} 
\end{figure*}

\clearpage
\subsubsection{Z-types}

\begin{figure*}[!htbp]
    \centering
    \includegraphics[width=0.9\textwidth]{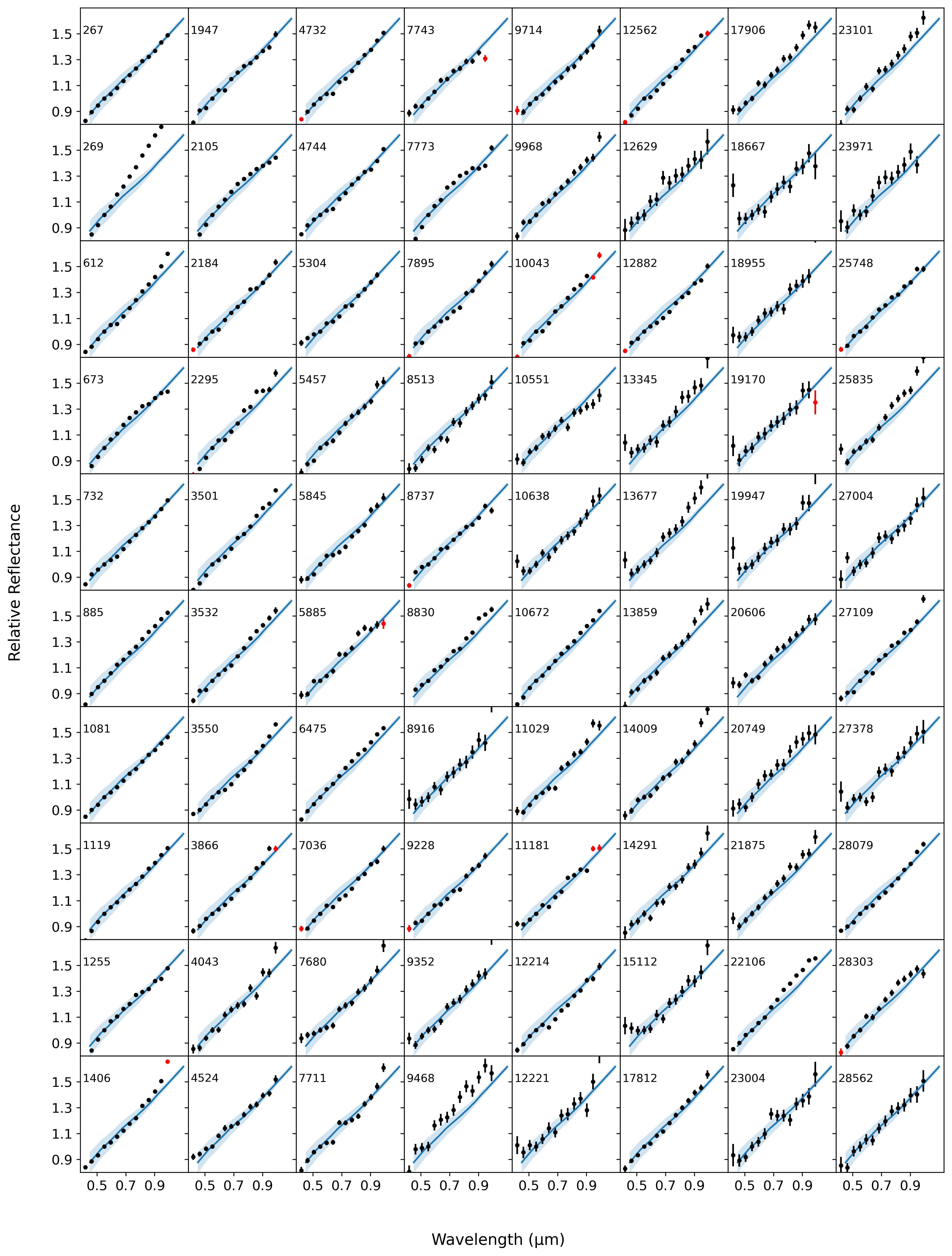}
    \caption{Main belt Z-types spectrophotometry from Gaia DR3 catalog (the black dots), with superimposed the best fit class from Mahlke taxonomy (blue line) \citep{Mahlke_2022}. The red dots represent data to be considered with caution (flagged 1 in DR3 catalog).} 
\end{figure*}

\begin{figure*}[!htbp]
    \centering
    \includegraphics[width=0.9\textwidth]{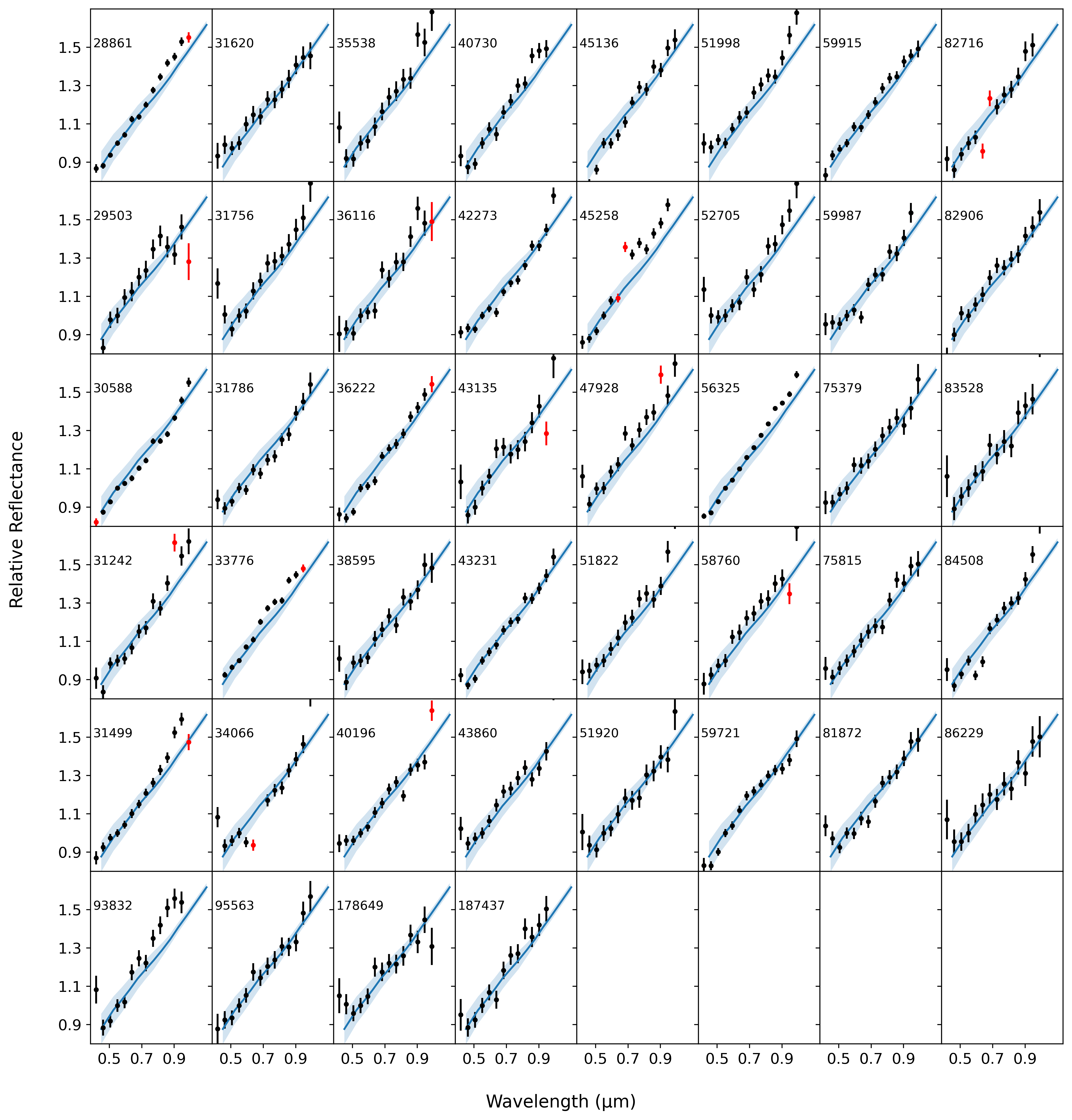}
    \caption{Main belt Z-types spectrophotometry from Gaia DR3 catalog (the black dots), with superimposed the best fit class from Mahlke taxonomy (blue line) \citep{Mahlke_2022}. The red dots represent data to be considered with caution (flagged 1 in DR3 catalog).}  
\end{figure*}
\clearpage
\subsection{Cybele}

\begin{figure*}[!htpb]
    \centering
    \includegraphics[width=0.9\textwidth]{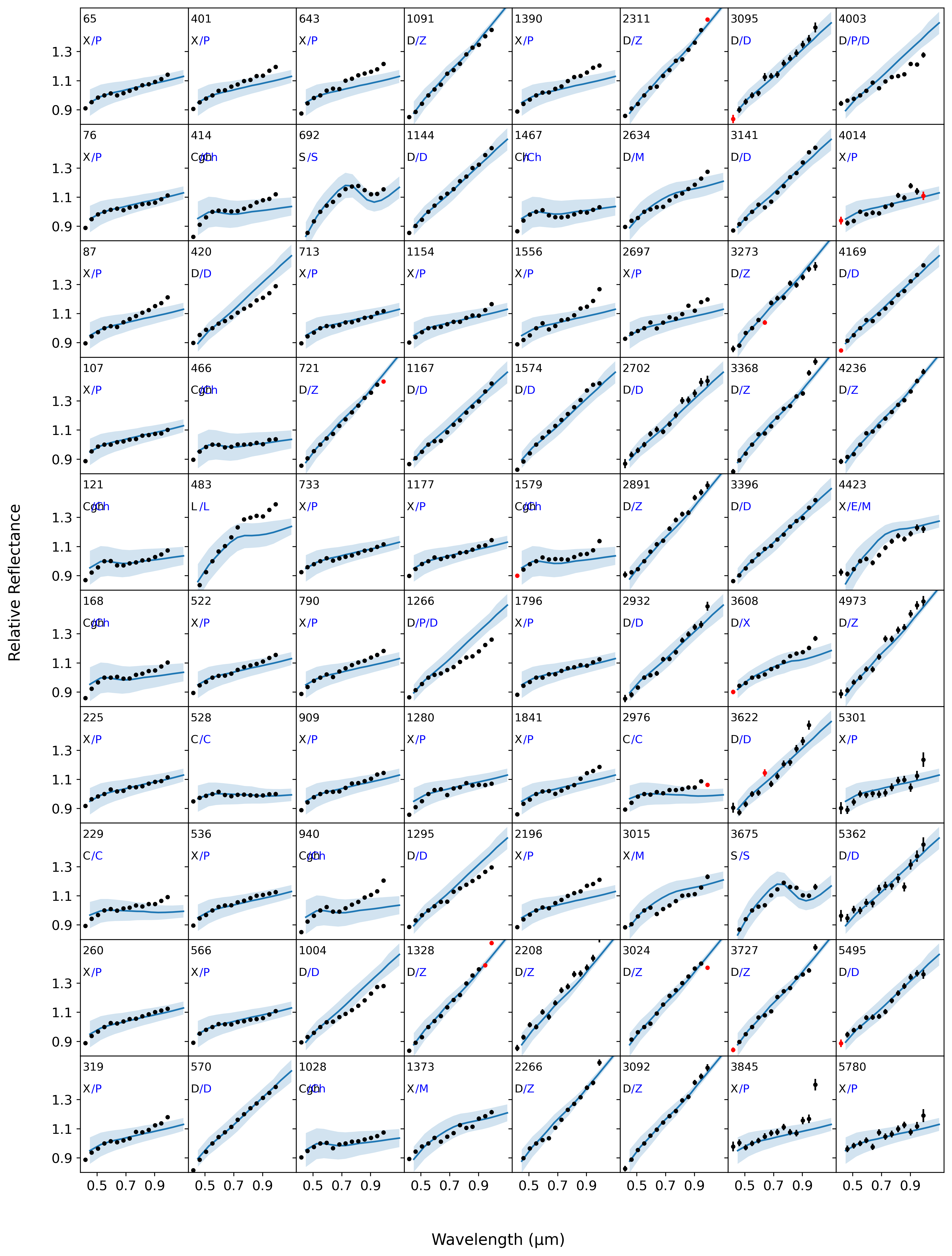}
    \caption{Cybele asteroids spectrophotometry from Gaia DR3 catalog (the black dots), with superimposed the best fit class from Mahlke taxonomy (blue line) \citep{Mahlke_2022}. The red dots represent data to be considered with caution (flagged 1 in DR3 catalog). The letters represent the best fit classes in Bus-DeMeo \citep{demeo_tax} (in black) and Mahlke (in blue) taxonomies.} 
\end{figure*}

\begin{figure*}[!htpb]
    \centering
    \includegraphics[width=0.9\textwidth]{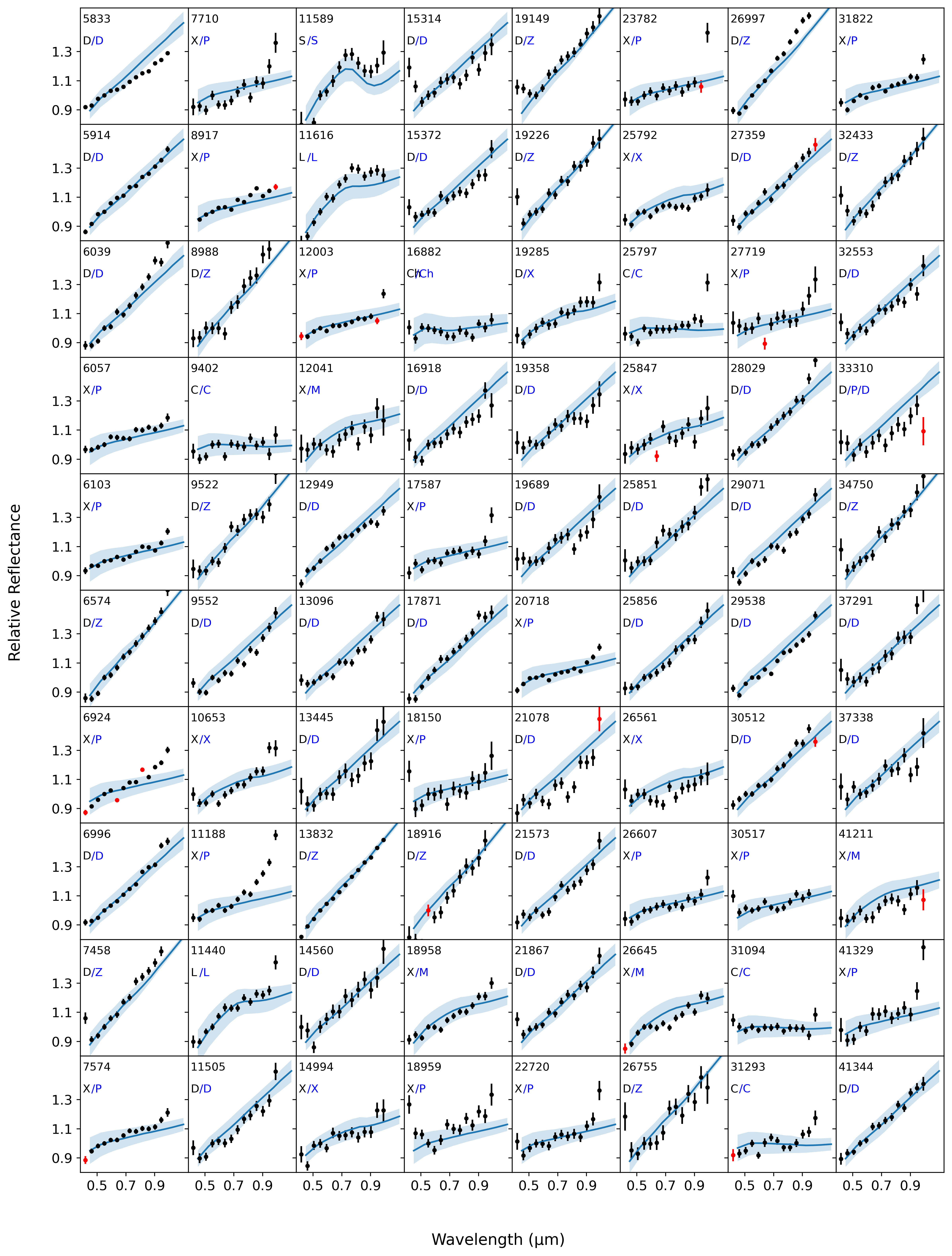}
    \caption{Cybele asteroids spectrophotometry from Gaia DR3 catalog (the black dots), with superimposed the best fit class from Mahlke taxonomy (blue line) \citep{Mahlke_2022}. The red dots represent data to be considered with caution (flagged 1 in DR3 catalog). The letters represent the best fit classes in Bus-DeMeo \citep{demeo_tax} (in black) and Mahlke (in blue) taxonomies.} 
\end{figure*}

\begin{figure*}[!htpb]
    \centering
    \includegraphics[width=0.9\textwidth]{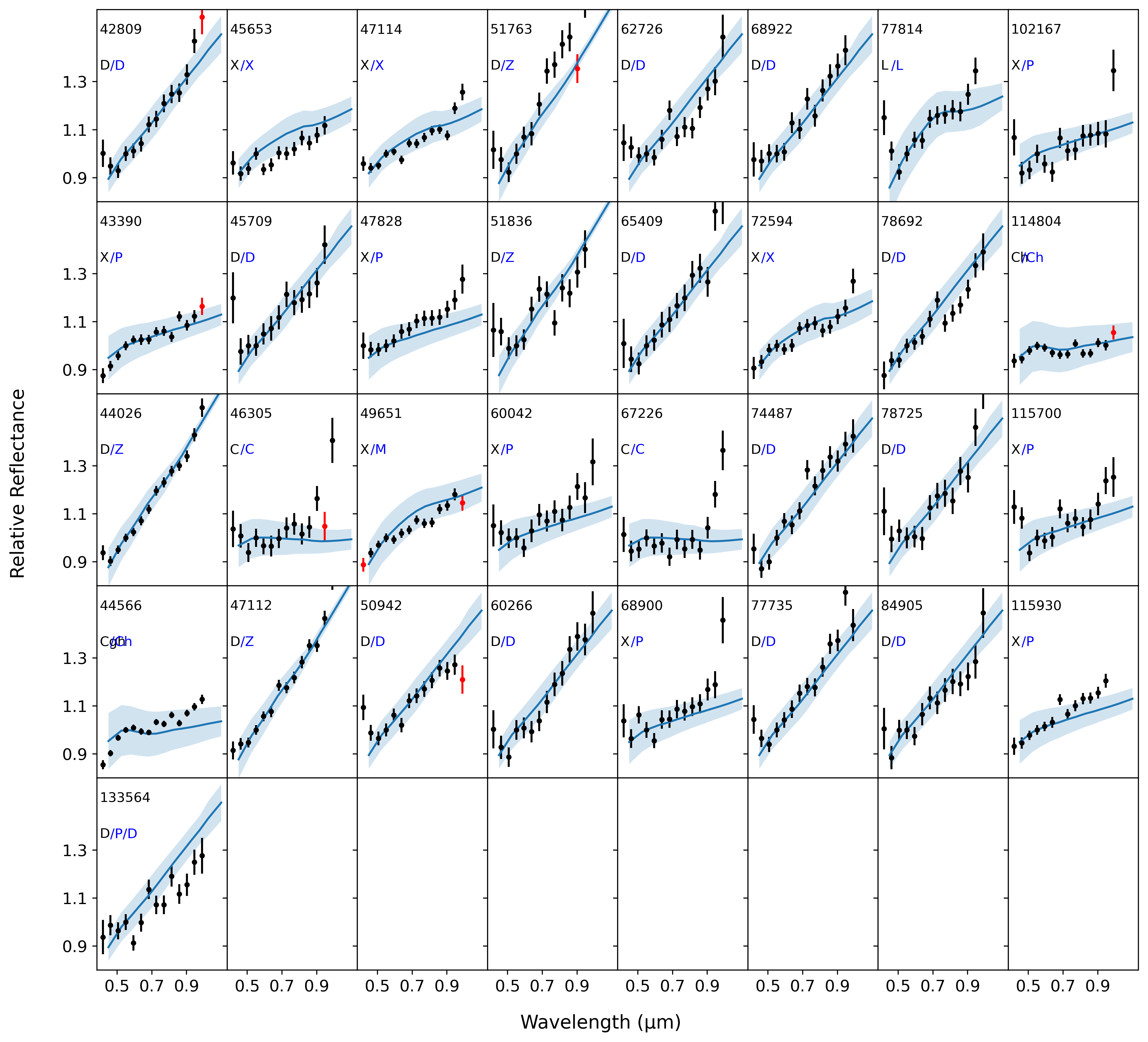}
    \caption{Cybele asteroids spectrophotometry from Gaia DR3 catalog (the black dots), with superimposed the best fit class from Mahlke taxonomy (blue line) \citep{Mahlke_2022}. The red dots represent data to be considered with caution (flagged 1 in DR3 catalog). The letters represent the best fit classes in Bus-DeMeo \citep{demeo_tax} (in black) and Mahlke (in blue) taxonomies.} 
\end{figure*}

\clearpage
\subsection{Hilda}
\begin{figure*}[!htpb]
    \centering
    \includegraphics[width=0.9\textwidth]{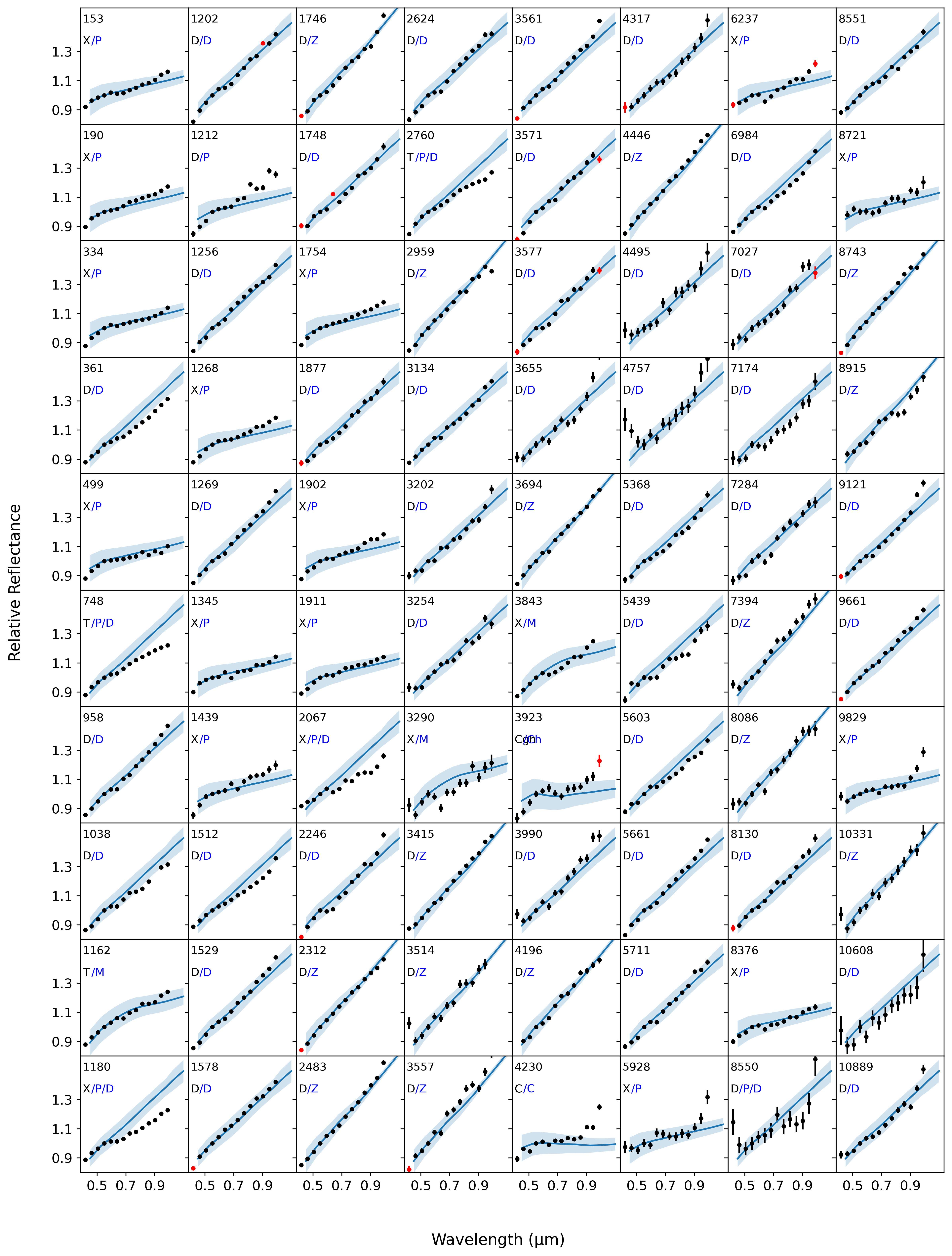}
    \caption{Hilda asteroids spectrophotometry from Gaia DR3 catalog (the black dots), with superimposed the best fit class from Mahlke taxonomy (blue line) \citep{Mahlke_2022}. The red dots represent data to be considered with caution (flagged 1 in DR3 catalog). The letters represent the best fit classes in Bus-DeMeo \citep{demeo_tax} (in black) and Mahlke (in blue) taxonomies.} 
\end{figure*}

\begin{figure*}[!htpb]
    \centering
    \includegraphics[width=0.9\textwidth]{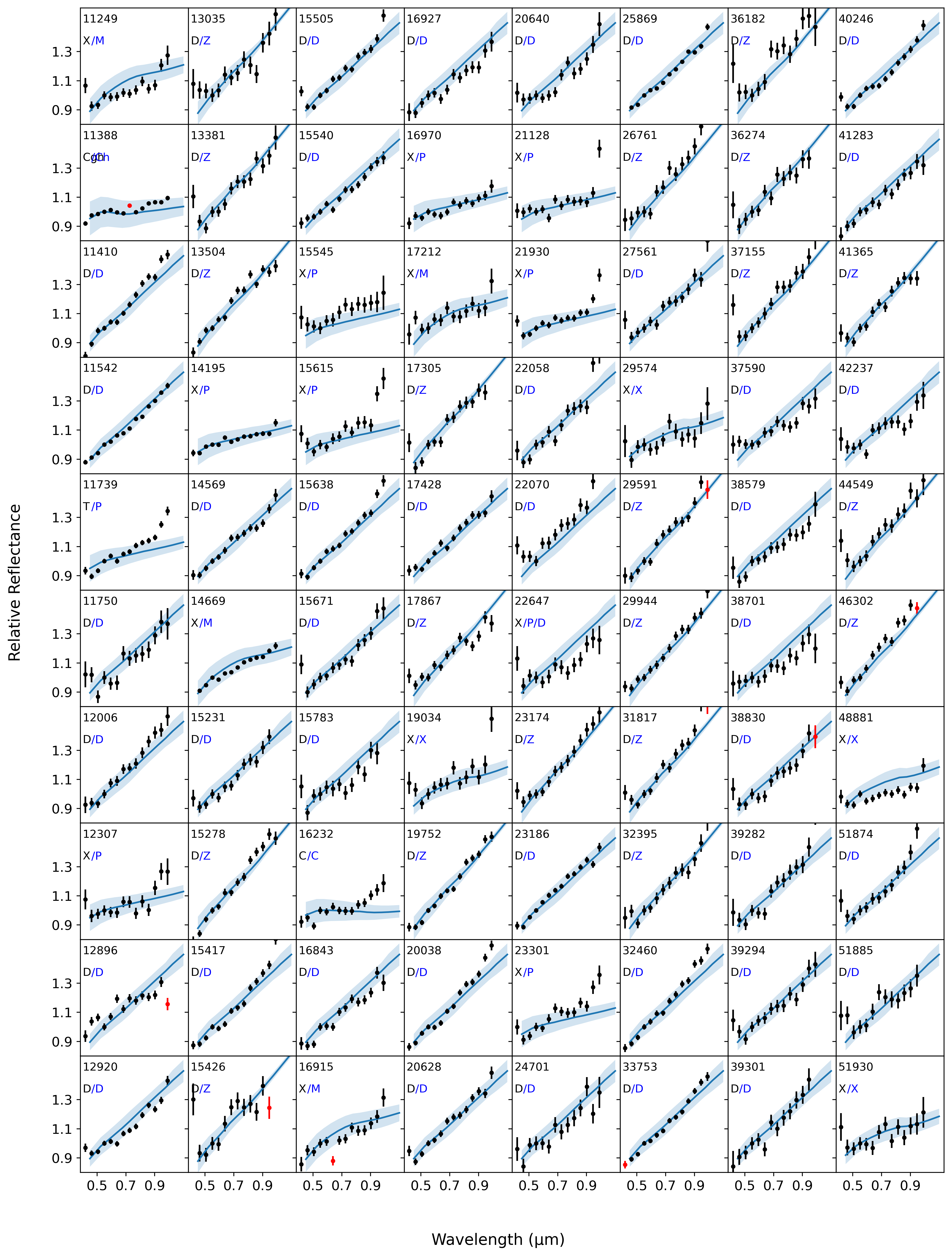}
    \caption{Hilda asteroids spectrophotometry from Gaia DR3 catalog (the black dots), with superimposed the best fit class from Mahlke taxonomy (blue line) \citep{Mahlke_2022}. The red dots represent data to be considered with caution (flagged 1 in DR3 catalog). The letters represent the best fit classes in Bus-DeMeo \citep{demeo_tax} (in black) and Mahlke (in blue) taxonomies.} 
\end{figure*}

\begin{figure*}[!htpb]
    \centering
    \includegraphics[width=0.9\textwidth]{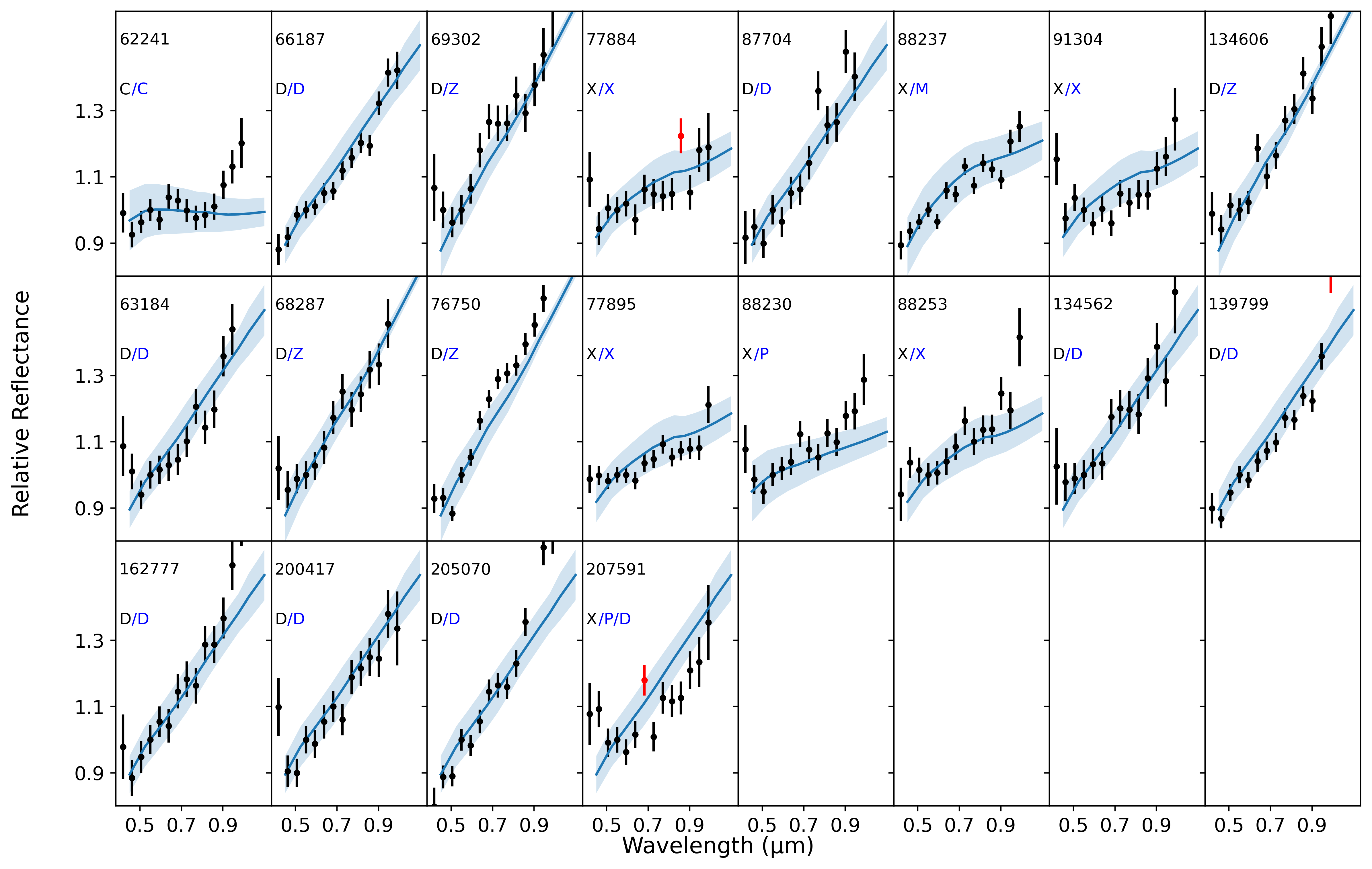}
    \caption{Hilda asteroids spectrophotometry from Gaia DR3 catalog (the black dots), with superimposed the best fit class from Mahlke taxonomy (blue line) \citep{Mahlke_2022}. The red dots represent data to be considered with caution (flagged 1 in DR3 catalog). The letters represent the best fit classes in Bus-DeMeo \citep{demeo_tax} (in black) and Mahlke (in blue) taxonomies.} 
\end{figure*}
\clearpage
\section{Additional table}
\captionsetup{type=table}
\caption{Investigated asteroids: $T_1$ and $T_2$ are the Bus-DeMeo \protect\citep{demeo_tax}, 
and Mahlke taxonomies \protect\citep{Mahlke_2022}, respectively. 
Spectral references: (1) \protect\citet{galluccio:hal-04225764}, 
(2) \protect\citet{oriel_2024_red}.}
\label{tab:appendix}
\begin{longtable}{llllllllll}
\toprule
Number & D (km) & Slope & $p_v$ & i (°) & $T_1$ & $T_2$ & SNR & Family & References \\
\midrule
\endfirsthead
\toprule
Number & D (km) & Slope & $p_v$ & i (°) & $T_1$ & $T_2$ & SNR & Family & References   \\
\midrule
\endhead
\midrule
\multicolumn{10}{r}{Continued on next page} \\
\midrule
\endfoot
\bottomrule
\endlastfoot
46 & 120.82 $\pm$ 1.25 & 3.091 $\pm$ 0.100 & 0.052 $\pm$ 0.002 & 2.4 & X & P & 539.7 &   & 1 \\
65 & 263.24 $\pm$ 2.78 & 1.905 $\pm$ 0.050 & 0.064 $\pm$ 0.002 & 3.6 & X & P & 1109.1 &   & 1 \\
72 & 79.64 $\pm$ 0.60 & 8.704 $\pm$ 0.094 & 0.072 $\pm$ 0.003 & 5.4 & D & D & 613.6 &   & 1 \\
76 & 155.51 $\pm$ 1.00 & 1.525 $\pm$ 0.084 & 0.039 $\pm$ 0.001 & 2.1 & X & P & 633.9 &   & 1 \\
81 & 119.96 $\pm$ 0.55 & 2.391 $\pm$ 0.091 & 0.047 $\pm$ 0.001 & 7.8 & X & P & 595.5 & Terpsichore & 1 \\
87 & 251.77 $\pm$ 1.70 & 3.778 $\pm$ 0.046 & 0.043 $\pm$ 0.002 & 10.9 & X & P & 1230.0 & Sylvia & 1 \\
107 & 215.33 $\pm$ 2.31 & 1.965 $\pm$ 0.042 & 0.058 $\pm$ 0.003 & 10.0 & X & P & 1266.5 & Sylvia & 1 \\
121 & 193.77 $\pm$ 1.80 & -0.727 $\pm$ 0.060 & 0.047 $\pm$ 0.001 & 7.6 & Cgh & Ch & 885.9 &   & 1 \\
\end{longtable}

\end{document}